\documentclass[pdflatex,sn-mathphys-num]{sn-jnl}

\usepackage{graphicx}%
\usepackage{multirow}%
\usepackage{amsmath,amssymb,amsfonts}%
\usepackage{amsthm}%
\usepackage{mathrsfs}%
\usepackage{xcolor}%
\usepackage[version=4]{mhchem} 
\usepackage{textcomp}%
\usepackage{manyfoot}%
\usepackage{booktabs}%
\usepackage{subcaption}
\usepackage{cleveref}
\usepackage{booktabs, subcaption}  
\usepackage{xparse}   

\usepackage{pifont}     
\usepackage{xcolor}     
\newlength{\goldpad}     
\newlength{\nuclidepad}  

\usepackage{algorithm}%
\usepackage{algorithmicx}%
\usepackage{algpseudocode}%
\usepackage{listings}%

\usepackage{booktabs}   
\usepackage{tabularx}   
\usepackage{siunitx}    

\usepackage{titlesec}   
\usepackage{etoolbox}  
\usepackage{chngcntr}
\usepackage[T1]{fontenc}

\pretocmd{\appendix}{%
  \titleformat{\section}                             
    {\normalfont\Large\bfseries}                     
    {Appendix~\thesection:}{1em}{}                   
}{}{}  

\theoremstyle{thmstyleone}%
\theoremstyle{thmstyletwo}%

\theoremstyle{thmstylethree}%

\begin{document}

\title[Article Title]{Scalable Chrysopoeia via $(n, 2n)$ Reactions Driven by Deuterium-Tritium Fusion Neutrons}


\author*[]{\fnm{Adam} \sur{Rutkowski*}}\email{adam@marathonfusion.com}

\author[]{\fnm{Jake} \sur{Harter}}

\author[]{\fnm{Jason} \sur{Parisi}}

\affil[]{\orgname{Marathon Fusion}, \orgaddress{\street{150 Mississippi}, \city{San Francisco}, \postcode{94107}, \state{CA}, \country{USA}}}


\abstract{ \unboldmath A scalable approach for chrysopoeia---the transmutation of base metals into gold---has been pursued for millennia. While there have been small-scale demonstrations in particle accelerators and proposals involving thermal neutron capture, no economically attractive approach has yet been identified. We show a new scalable method to synthesize stable gold (\ce{^{197}Au}) from the abundant mercury isotope $\ce{^{198}Hg}$ using $(n, 2n)$ reactions in a specialized neutron multiplier layer of a fusion blanket. Reactions are driven by fast $14\, \mathrm{MeV}$ neutrons provided by a deuterium-tritium fusion plasma, which are uniquely capable of enabling the desired reaction pathway at scale. Crucially, the scheme identified here does not negatively impact electricity production, and is also compatible with the challenging tritium breeding requirements of fusion power plant design because $(n, 2n)$ reactions of $\ce{^{198}Hg}$ drive both transmutation and neutron multiplication. Using neutronics simulations, we demonstrate a tokamak with a blanket configuration that can produce \ce{^{197}Au} at a rate of about \SI{2}{t/GW_{th}/yr}. Implementation of this concept allows fusion power plants to double the revenue generated by the system, dramatically enhancing the economic viability of fusion energy. }



\maketitle

\section{Introduction}\label{sec1}

Significant technical hurdles must be overcome to achieve net energy production in fusion devices \cite{community2020,Batani2023future,Fasoli2023,Meyer2024}. Even more important, deployment of fusion energy at the scale required to achieve energy abundance and solve climate change requires the approach to be economically competitive \cite{Lindley2023}. This is an especially difficult task for a technology as complex as fusion where the end product is electricity, a low-value commodity that can be produced from many other sources of primary energy.

The fusion economics challenge has motivated further investigation of high-value products that can be made with fusion neutrons through transmutation. Previous efforts in fusion have aimed to identify higher-value applications of fusion neutrons in early markets, including neutron imaging, medical isotope production, and fission waste burning \cite{engholm1986radioisotope,shine_neutron_imaging,Li2023_FusionIsotopes, Murgo2023_MAtransmutation}. While the value for neutrons used in these ways is much larger than the value of electricity that would otherwise be produced by these neutrons, there is a limited market size for these applications, and the construction of one or a few fusion power plants at large (gigawatt of thermal power, \si{GW_{th}}) scale would saturate the markets for neutron imaging and medical isotope production, which are \$100\textendash$200\mathrm{M/yr}$ and around \$5B/yr, respectively \cite{Honney2023FusionNeutrons, WNA2025RadioisotopesMedicine}. An ideal product of fusion transmutation would substantially supplement the value of electricity generated by the device while serving a market sufficiently large to subsidize the deployment of TW-scale fusion energy.

In this work, we show how a particular category of neutron-driven reactions---the $(n,2n)$ multiplication reactions required in all deuterium-tritium (D-T) fusion blankets---enables scalable transmutation of mercury into stable gold while still meeting the broader requirements of the fusion system. We show with neutronics simulations that a system optimized to use these reactions for transmutation can produce \ce{^{197}Au} at a rate of more than \SI{2}{t/GW_{th}/yr}, which more than doubles the value of outputs from the fusion system.  This approach is a solution to the longstanding problem of chrysopoeia, the transmutation of base metals into gold.  Because of the sizable gold market, this approach can subsidize the deployment of terawatts of fusion power without saturation.

The layout of this work is as follows.
In \Cref{sec:neutronics_consid} we introduce the basic neutronics considerations relevant to transmutation using different neutron sources.  In \Cref{sec2}, we introduce the general approach to chrysopoeia in a fusion power plant. In \Cref{sec3}, we present specific examples of fusion systems designed for chrysopoeia and compare their properties. We discuss the key challenges to implementation of these systems in \Cref{sec4}, and we conclude in \Cref{sec5}.

\section{Neutronics Considerations} \label{sec:neutronics_consid}

In this work, we focus on the use of neutrons to drive nuclear reactions.  Because they lack an electrical charge, neutrons are not deflected by the Coulomb force from the charged target nucleus and so can in general interact with it more directly than charged particles can.  The neutron-driven reactions most relevant to this work are the $(n,\gamma)$ neutron capture reaction and the $(n,2n)$ neutron multiplication reaction.  The $(n,\gamma)$ reaction results in capture of a neutron and emission of a photon, increasing the atomic mass of the target nucleus by one while leaving the atomic number unchanged. The $(n,2n)$ reaction results in the opposite effect, releasing on net one neutron from the nucleus. The $(n, \gamma)$ reaction has highest cross section for low energy neutrons, while the $(n,2n)$ reaction is accessible only to high energy neutrons.  We focus on $(n,2n)$ reactions for transmutations of interest because the reaction is critical to the neutron economy of the system, a constraint we will discuss further in Section~\ref{sec2}.  

In this section we discuss some of the differences between fission and fusion neutron sources, and show that fusion is uniquely capable of providing large quantities of high energy neutrons useful for driving new and valuable transmutation pathways.  




\subsection{Fission Neutrons}
We first consider the neutron properties of a fission system that are relevant for transmutation. 
Thermal‐neutron induced fission of $^{235}$U or $^{239}$Pu results in:
\[
  \begin{aligned}
    \ce{^{235}_{92}\mathrm{U}} + \ce{^1_0n}
      &\longrightarrow \text{Fission fragments} + \nu_\mathrm{U}\,\ce{^1_0n},\\
    \ce{^{239}_{94}\mathrm{Pu}} + \ce{^1_0n}
      &\longrightarrow \text{Fission fragments} + \nu_{\mathrm{Pu}}\,\ce{^1_0n}\,.    
  \end{aligned}
\]
with $\nu$ providing the average number of neutrons per fission, and a neutron energy spectrum with neutron rate peaking around $0.75\,\mathrm{MeV}$, with very small emission near 7--8$\,\mathrm{MeV}$~\cite{nereson_fission_1952}. 

At least two issues arise using neutrons sourced from fission systems for transmutation: first, the high cost of excess neutrons, and second, the low energy of the emitted neutrons. In fission systems, neutron economy is a critically important aspect of power plant design and most neutrons are needed to drive downstream reactions to maintain operation in steady state. Some neutrons can be harvested from the edges of a fission device, or specialized test reactors can be built for samples to be inserted into high neutron flux regions. For example, the High Flux Isotope Reactor (HFIR) at ORNL achieves relatively high neutron density through the use of $\ce{^{235}U}$ fuel highly enriched to $\sim93 \mathrm{wt}\%$ \cite{Ilas2016_HFIR_KeyMetrics}. Alternatively, specialized designs such as fast breeder reactors can produce significant amounts of excess neutrons that can be used to breed additional fuel, with the Superfénix power plant designed to achieve an excess breeding ratio (defined as the breeding ratio minus one) of $0.2$ \cite{Vendryes1985}. 

However, while these modifications can provide an excess of neutrons, they are expensive and still provide only low energy neutrons. This significantly limits the range of downstream reactions, and for the reactions that we propose here large quantities of neutrons above $\sim 9\,\mathrm{MeV}$ are required, as we will explain shortly. This neutron energy issue is intrinsic to the nature of fission reactions and cannot be circumvented through design.

\subsection{Deuterium-Tritium Fusion Neutrons}
The D-T fusion reaction \cite{huba2004nrl}:
\begin{equation}\label{eq:D-T}
  \ce{^2_1H + ^3_1H}
  \;\longrightarrow\;
  {}^{4}_{2}\mathrm{He}\,\bigl(E_{\alpha}\approx3.5\,\mathrm{MeV}\bigr)
  \;+\;
  {}^{1}_{0}n\,\bigl(E_{n}\approx14.1\,\mathrm{MeV}\bigr)
\end{equation}
has the special property of producing neutrons with \SI{14.1}{MeV} of energy.  This is due to the large difference in binding energy per nucleon between the reactants D and T and the product $\ce{^4He}$, as well as the fact that due to conservation of momentum and the large difference in mass between the products of the reaction, the neutron carries 80\% of the energy released.  In contrast, the deuterium-deuterium (D-D) reaction produces a neutron with only \SI{2.45}{MeV} of energy \cite{huba2004nrl}. 

  To remove a neutron from a target nucleus, an incoming neutron must interact with this nucleus, and then impart enough energy to remove two neutrons from the system---both the incident neutron and the additional neutron. This sets the threshold for the endothermic $(n,2n)$ reaction, which for almost all nuclei is in the range of 6--9\,MeV.  The ability to access this $(n,2n)$ reaction is thus a special property of the D-T fusion neutron, enabling a wide range of additional transmutation reactions beyond what would be accessible only with neutron capture reactions.  
  
\begin{figure}[htbp]
  \centering
  \includegraphics[width=0.99\textwidth]{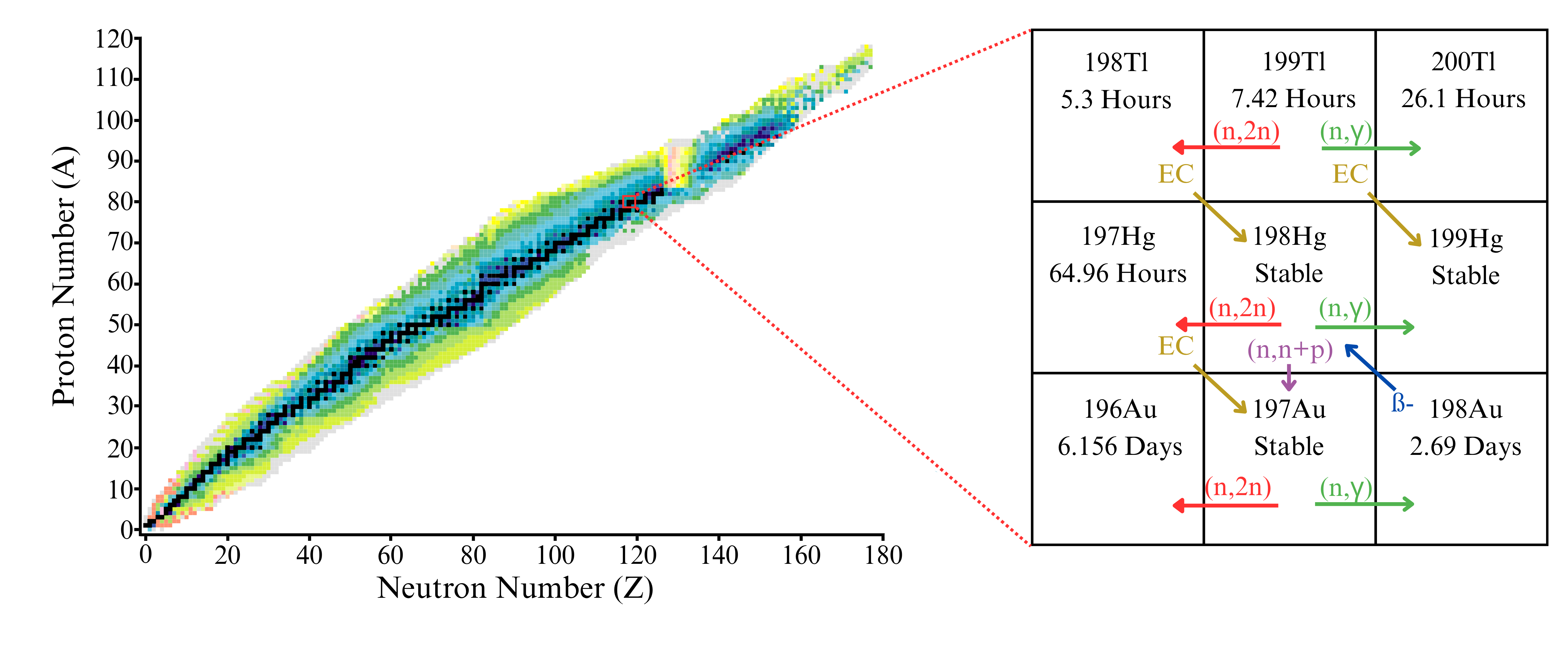}
   \caption{Table of nuclides (from NuDat.\cite{nndc_nudat3}) with a zoomed view of the region of interest.
   Arrows show how neutron-driven reactions and radioactive decay direct transmutation.
   EC stands for electron capture.
   } 
  \label{fig:nuclidechart}
\end{figure}

  In Figure~\ref{fig:nuclidechart}, the table of nuclides is shown alongside a zoomed-in view of the portion of the table of most interest for the desired transmutation in this work.  Different neutron driven reactions and decay processes are shown and illustrate how different pathways and products can be accessed with new reactions. 

While opening up a new set of reaction pathways is promising, any approach to use these reactions has to account for the unique constraints of the D-T system, in particular due to the requirements for the costly tritium needed as fuel.  The following section describes the constraints inherent to fusion blanket design and demonstrates the possibility of designing a blanket that simultaneously meets the fuel cycle requirements of D-T fusion and achieves economically valuable production of gold.

\section{Transmutation by $(n,2n)$ in Fusion Systems}\label{sec2}

Two key observations underlie the scheme proposed here. First, a certain set of neutron-driven reactions aside from the tritium-producing reaction are \textit{always required} in a fusion system to close the fuel cycle and achieve tritium self-sufficiency. Second, if properly selected, these reactions can be used to produce high-value outputs in a fusion device without sacrificing key performance parameters.  

\subsection{Fuel Cycle Constraints}

The nature of the D-T fuel cycle requires that the neutron released by this reaction be used to breed tritium to sustain the reaction.  

While some tritium can be produced from neutron capture on deuterium in heavy water moderated power plants at a rate of $\sim 130\,\mathrm{g/yr/GW}$, this is far from sufficient to produce the required tritium for full-scale fusion power plants, which burn $\mathrm{56\,kg/GW_{th}}$ per full power year \cite{abdou_physics_2021}.
The only way to produce the required tritium at the scale needed for D-T power plant operation is through the system itself---namely by using the D-T fusion neutron to create replacement fuel, a reaction which occurs primarily through a reaction with a thermal neutron on \ce{^6Li} \cite{huba2004nrl}:
\begin{equation}\label{eq:li6nt}
\ce{^6_{3}Li + ^{1}_{0}$n$ -> ^{3}_{1}H + ^{4}_{2}He}
\quad(Q = 4.8\ \mathrm{MeV})
\end{equation}
where $Q$ denotes the energy released in the reaction.
Any proposed use of fusion neutrons must ensure that tritium breeding is not adversely impacted.
The tritium breeding ratio (TBR) in a fusion device is defined as the ratio of tritium produced in the blanket to tritium consumed in the fusion reaction.
A minimal condition for scalable fusion is $\mathrm{TBR}>1$, with the required excess TBR determined by radioactive decay losses, nonradioactive losses in the system, and the desired scaling rate for fusion devices (since the tritium inventory for subsequent power plants must be provided by tritium generated from earlier devices). In most fusion systems, a TBR around 1.1--1.2 is targeted, and many blanket designs have been modeled to achieve this range through Monte~Carlo neutronics simulations \cite{sorbom_arc_2015,najmabadi_aries-at_2006,liu_conceptual_2014,hernandez_new_2017}.
It should be noted that because of the difficulty of any source besides fusion to provide the required tritium for supplying fusion at scale, the need for $\mathrm{TBR}>1$ is an inflexible requirement with current technologies.  

However, the absolute best case that could be achieved through the \ce{^6Li} reaction~\eqref{eq:li6nt} and a thermal neutron is $\mathrm{TBR}=1$. This is because each fusion reaction consumes one tritium nucleus and produces a single neutron, and reaction~\eqref{eq:li6nt} produces only a single triton per neutron.  In practice, this best-case scenario is unattainable due to neutron losses to other reactions, including neutron capture in structural materials required in the system.  

To solve this problem, all D-T power plants rely on some form of neutron multiplication to achieve tritium self-sufficiency.
Almost all proposed power plant designs rely on the $(n,2n)$ neutron multiplication reaction in $\mathrm{Be}$, $\mathrm{Pb}$, or $^7\mathrm{Li}$ for this purpose \cite{Pettinari2022ARC,Clark2025InfinityTwo,Ogando2024HYLIFEIII}. In the case of $^7\mathrm{Li}$, there is also a $\ce{^7Li}(n,n\alpha)\ce{T}$ reaction that contributes to tritium production more than the $(n,2n)$ reaction. This reaction preserves the neutron and produces a tritium, but this reaction also requires high energy neutrons ($\gtrsim 4\,\mathrm{MeV}$) to proceed, and so the behavior is similar to that of the $(n, 2n)$ reaction and we will discuss this reaction in the same terms as $(n,2n)$ reactions in this work.  Aside from purpose-designed multiplier layers, we note that structural materials also contribute to the total $(n,2n)$ rate, though this function for the structural layer is typically not optimized as a design feature in studies. 

These multiplier materials can be implemented as independent layers integrated into the blanket radial build, or chemically integrated in the blanket fluid itself.  The three options $\mathrm{Be}$, $\mathrm{Pb}$, or $^7\mathrm{Li}$ correspond directly to the most common choices for liquid breeder blanket working fluids: fluorine-lithium-beryllium (FLiBe) molten salt, lithium-lead (LiPb) eutectic, and liquid lithium.  A more recent work has also considered chloride salts, and in particular noted the benefit of salts enriched in $^{37}\mathrm{Cl}$ to improve tritium breeding properties, including the observation that $^{37}\mathrm{Cl}$ has a higher $(n,2n)$ cross section than $^{35}\mathrm{Cl}$ \cite{Bohm2023ClSalt}.   

In sharp contrast with the previous comments on the high cost for thermal neutron-driven reactions, the absolute necessity for large numbers of $(n,2n)$ to occur in a D-T fusion blanket means that if the correct conditions can be met for a multiplier material, the products of the reactions driven in this material when implemented in the multiplier layer of a fusion blanket are ``free'' from the perspective of the system design. 

\subsection{Requirements for Multiplier Materials} \label{sec:multiplierrequirements}

We propose using $(n, 2n)$ reactions on neutron multiplier materials carefully selected to produce a high-value product as a result of the $(n, 2n)$ reaction and any subsequent decays.  
To be a feasible candidate for use in a specialized multiplier layer, the following conditions should be met:
\begin{enumerate}
  \item The material must have sufficient cross section for the $(n, 2n)$ reaction to contribute substantially to TBR.
  \item The $(n, \gamma)$ neutron capture cross section must not be too large, since this represents a parasitic loss of neutrons to undesired reactions.  This reaction also produces gamma rays that will heat the superconducting magnets in magnetic confinement devices.
  \item The feedstock isotope must be sufficiently abundant and low‐cost that there are minimal supply challenges.
  \item An ideal feedstock will not produce long‐lived radioactive waste, or can undergo isotope separation before use to limit the amount of long‐lived waste produced.
  \item The product to be made from the feedstock should have only limited long‐lived radioactive isotopes generated, since downstream use in many applications would require isotopic separation of these impurities at very low concentration.   
    \item Because any product will be generated at relatively low concentrations in the multiplier layer, it is important that the eventual product \emph{not be the direct product of the $(n, 2n)$ reaction, but a result of a radioactive decay of the product of the $(n,2n)$ reaction which makes a new element}. This means that the product can be separated chemically rather than isotopically from the multiplier material.  This property also means that different and potentially much more valuable elements can be made from low‐cost, abundant feedstock elements.
  \item For the proposed product to provide additional value for \textit{many} future fusion devices, the market for this product should be sufficiently large that it will not be saturated by only a few fusion devices making the desired material.
\end{enumerate}

Feedstocks and products must also be capable of integration with the numerous requirements inherent to the environment of the fusion blanket. Specifically they must be compatible with structural materials at high temperature, and the thermophysical properties (melting point, vapor pressure, and tritium solubility, to name the most critical) must be such that a self‐consistent blanket design can be engineered around the material.

A simple heuristic calculation can aid in understanding the feasibility of this approach to generate economically relevant quantities of material if the right feedstock material can be identified. In order to develop intuition, in this discussion as well as later in this work we will neglect the contribution of blanket energy multiplication on total power output since this effect depends on the specifics of each blanket design. As shown in the simplified blanket material scans in \cite{sawan_physics_2006}, even between natural enrichment FLiBe and $90\%$ enriched LiPb breeder material, two blanket materials with markedly different neutronic properties, there is only a roughly $3\%$ difference in energy multiplication factor, justifying our choice to neglect changes in this parameter in our analysis. 

Under this simplification, there are $\sim 1.12\times 10^{28}$ fusion reactions per year in a \SI{1}{GW_{th}} fusion device. Assuming neutron multiplication is dominated by $(n,2n)$ reactions, in order to achieve a $\mathrm{TBR}=1.2$, at an absolute minimum $20\%$ of all fusion reactions must have a corresponding $(n, 2n)$ reaction in the blanket; as a less conservative value, it is known that simplified blanket configurations (\SI{2}{m} thick, no structure, natural Li) can achieve as high as $\mathrm{TBR}\sim1.85$ \cite{sawan_physics_2006}, in which case at least 85\% of fusion reactions must have a corresponding multiplication reaction. This range corresponds to $3.7\times10^3$\textendash$1.6\times10^4 \ \mathrm{mol/yr}$, or for a product with a mass of \SI{197}{amu}, $732-3114 \ \mathrm{kg/yr}$ of material production.  

To be worth deploying in a fusion power plant, we can make the heuristic assumption that implementation of a specialized blanket layer must increase the value produced by the device by at least $20\%$.  A \SI{1}{GW_{th}}  power plant operating continuously with 40\% conversion efficiency selling power at $\$50/\mathrm{MWh_e}$ (a price target identified in prior literature \cite{Handley2021EarlyFusionMarkets}), would generate $\$175\mathrm{M/yr}$ in revenue.
To meet the target of 20\% of revenue, or $\$35\mathrm{M/yr}$, generated by reaction products,
the value of the product would need to be in the range of $\$11\mathrm{k}-\$48\mathrm{k}\mathrm{/kg}$.  

\subsection{Feedstock and Product Materials}

Given the long list of requirements listed above, it is surprising that there is a feedstock isotope, reaction pathway, and product material that meets both the technical and market requirements described above.  The feedstock material is the \ce{^{198}Hg} isotope of mercury, and the product is stable gold, $^{197}\mathrm{Au}$, which has a current market price $>\$\SI{100000}{/kg}$ \cite{LBMA_gold_price}, well above the price target proposed above. 

We specifically focus on the use of $(n,2n)$ multiplication reactions on \ce{^{198}Hg} to produce \ce{^{197}Hg} \cite{Luo2018HgN2N,Kasugai2001ActivationHg,Hg197m_MIRD_2025,KAERI_Hg197} and ultimately stable $^{197}\mathrm{Au}$, namely
\begin{align}
{}^{198}_{\,80}\mathrm{Hg} &\;(n,2n)\
{}^{197}_{\,80}\mathrm{Hg} + e^{-} \ \xrightarrow[\;T_{1/2}=64.1\ \mathrm{h}\;]{\varepsilon}\;
{}^{197}_{\,79}\mathrm{Au} + \nu_{e}
\\
\textrm{and}\quad \ce{^{198}_{80}Hg}\;(n,2n)\;&{}^{197\mathrm{m}}_{80}\mathrm{Hg}
   \;\xrightarrow[\;T_{1/2}=23.8\ \text{h}\;]{%
      \substack{94.7\%\,\mathrm{IT}\\5.3\%\,\varepsilon}}
   \begin{cases}
      \ce{^{197}_{80}Hg}\;+\;\gamma & (94.7\%)\\[4pt]
      \ce{^{197}_{79}Au}\;+\;\nu_{e} & (5.3\%)
   \end{cases}
\end{align}
where the cross sections for production of the ground state ${}^{197}_{80}\mathrm{Hg}$ and the isomer state ${}^{197\mathrm{m}}_{80}\mathrm{Hg}$ are similar \cite{Luo2018HgN2N}.

\begin{figure}[htbp]
  \centering
  \includegraphics[width=0.85\textwidth]{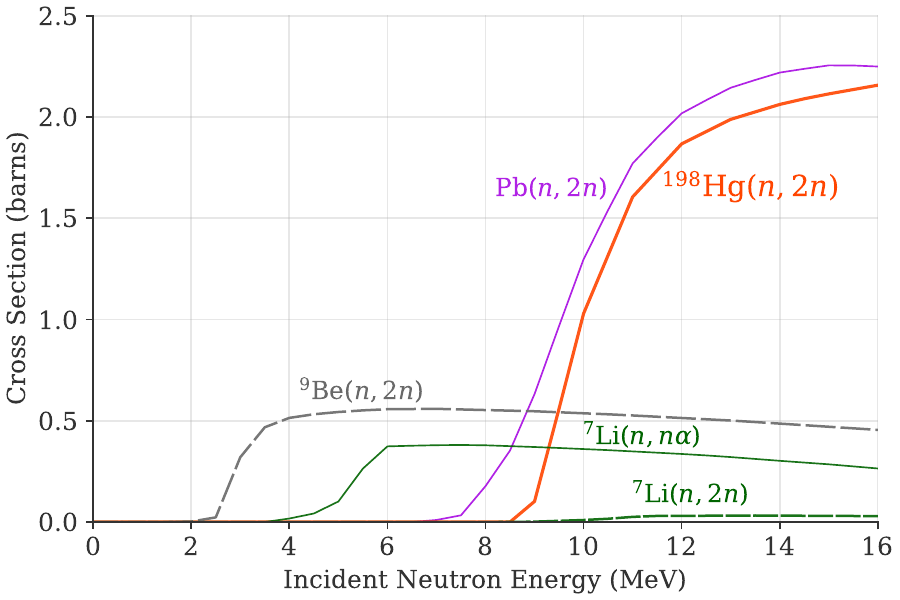}
  \caption{Cross sections for the $(n,2n)$ reaction for $\mathrm{Pb}$ and other common multiplier materials, as well as for \ce{^{198}Hg}, the multiplier proposed here. 
  The $(n,2n)$ cross section for lead is averaged across all naturally occurring isotopes with a weighting based on abundance. Data is from \cite{ENDFB8p1_Be9_n2n,ENDFB8p1_Pb_n2n, ENDFB8p1_Li7_n2n,ENDFB8p1_Hg198_n2n,Ge2020_CENDL32li7na}. }
  \label{fig:n2nxsections}
\end{figure}

While this product can also be generated through thermal neutron capture on \ce{^{196}Hg} as will be described shortly, the $(n,2n)$ reaction provides a far more appealing pathway for the economically viable production of gold than neutron capture, as the natural abundance of \ce{^{198}Hg} is $10\%$, 66 times larger than that of \ce{^{196}Hg}. However, this reaction has a threshold energy at $\sim$\SI{9}{MeV}, higher than the neutron energies available from either fission devices or D-D neutrons, but accessible to D-T fusion neutrons. 

 The cross section for the $(n,2n)$ reaction on \ce{^{198}Hg}, shown in Figure~\ref{fig:n2nxsections} alongside the averaged cross sections for other multiplier materials, is slightly smaller than that of natural $\mathrm{Pb}$ due to a higher threshold for the reaction, but the total cross section is comparable.  Critically, this cross section is sufficiently large to enable gold production and sufficient $(n,2n)$ reactions for tritium breeding, as we will show.  Mercury is cheap and abundant, and has good thermophysical properties that make its use in a blanket feasible.  

\subsection{Comparing Prior Approaches to Gold Production}

A number of pathways have been identified for producing gold from other elements through accelerators \cite{Aleklett1981}, but these approaches are not scalable due to extremely high cost for the infrastructure and inherently low production rates achievable.  Aside from the accelerator-based methods, it is well known that $^{197}\mathrm{Au}$, the only stable isotope of gold, can be produced through the decay of \ce{^{197}Hg} or $^{197}\mathrm{Pt}$. Because of practical considerations (abundance and cost), only \ce{^{197}Hg} is of interest. The isotope \ce{^{197}Hg} can be made using neutron capture by primarily thermal-spectrum neutrons on the naturally occurring isotope \ce{^{196}Hg} \cite{Brown2018_ENDFB8,KAERI_Hg197}:
\begin{equation}\label{eq:hg196}
{}^{196}_{80}\mathrm{Hg}\;(n,\gamma)\;{}^{197}_{80}\mathrm{Hg}
\; +\;e^{-}\xrightarrow[\;T_{1/2}=64.1\ \text{h}\;]{\varepsilon}\;
{}^{197}_{79}\mathrm{Au}\;\;+\;\nu_{e}.
\end{equation}
For this pathway, there are two key limitations: the aforementioned high cost of neutrons that can drive this reaction when provided by any existing neutron sources,
and the low natural abundance (0.15\%) of the \ce{^{196}Hg} isotope.

Even with an abundant source of low-cost neutrons, the extremely large quantities of feedstock mercury and the large amounts of isotope separation required would render a production scheme based on this pathway infeasible.  

We note that prior work has considered the possibility of transmuting mercury in a fusion blanket to produce gold. However, these works \cite{Bourque1988FAME, engholm1986radioisotope} predicted gold production of only $\sim \SI{200}{kg/GW_{th}/yr}$---an order of magnitude lower than the rate of $\sim \SI{2000}{kg/GW_{th}/yr}$ presented in this work---because the focus was on transmutation using thermal neutrons and multiplication with a separate fast neutron reaction on $\ce{^{9}Be}$. Specifically, \cite{Bourque1988FAME} states that detailed analysis is focused on $\ce{^{60}Co}$ production, ``with production rates for other isotopes generally deduced from the $\ce{^{60}Co}$ results''.  While not explicitly specified, it also appears that only natural enrichment mercury was assumed in this analysis. 

In contrast to their assumptions regarding thermal neutrons being the most important, the insight of this work is the desirability of transmutation and multiplication using a fast neutron reaction.
The key observations of this work are specifically that:
\begin{enumerate}
    \item The $(n,2n)$ cross section of mercury is sufficiently large to produce enough neutrons for downstream tritium production without the use of other multiplier materials. 
    \item Use of mercury enriched in \ce{^{198}Hg} can produce an economically relevant quantity of gold through $(n,2n)$ reactions, in contrast with $(n,\gamma)$ reactions on the low abundance \ce{^{196}Hg} isotope.
    \item Blanket designs optimized based on these observations can increase the yearly gold production per \si{GW_{th}} by a full order of magnitude from prior attempts.  
\end{enumerate}


While many other products can be made with the same methods described here, we focus in this work on the highest value case---gold production from \ce{^{198}Hg}---due to the economic significance of this material. Some other selected reactions and products are described in Appendix~\ref{sec:otherreactions}.

 The following sections will discuss how to implement blanket configurations that achieve large-scale production of gold, with some brief discussion of ancillary systems that can be implemented to enable this approach.  

\section{A Fusion System Designed for $^{197}\mathrm{Au}$ Production} \label{sec3}

By making use of the reactions described in the previous section, a fusion blanket can be designed to maximize gold production while achieving $\mathrm{TBR}>$1.1--1.2.  We specifically propose a two-layer system composed of an ``inner blanket'' and ``outer blanket'', where the two layers have different working fluids selected based on the different neutron energy spectra in each layer.  The inner blanket is designed to maximize the total number of $(n, 2n)$ reactions and consequently the generation of the desired product.  Because the neutron energy falls off exponentially over the blanket radial profile, the majority of all possible $(n, 2n)$ reactions can be achieved by having a relatively thin inner blanket layer which composes only a small fraction of the total blanket volume.
Since the mass of \ce{^{198}Hg} in the multiplier layer will drive the cost for mercury isotope separation to build a new fusion power plant and the availability of feedstock mercury may initially be constrained, it is important to minimize the total quantity of material required.  In order to further improve total blanket performance, we also propose operating with some fraction of lithium alloyed with the mercury layer to make the best possible use of the additional $(n, 2n)$-produced neutrons for breeding tritium.  Here we model this lithium as being enriched in \ce{^6Li}, which will maximize tritium breeding performance from thermal neutrons and minimize the number of high energy neutrons that perform $(n, 2n)$ reactions in the \ce{^7Li} instead of the feedstock \ce{^{198}Hg}.  

\begin{figure}[htbp]
  \centering
  \includegraphics[width=\textwidth]{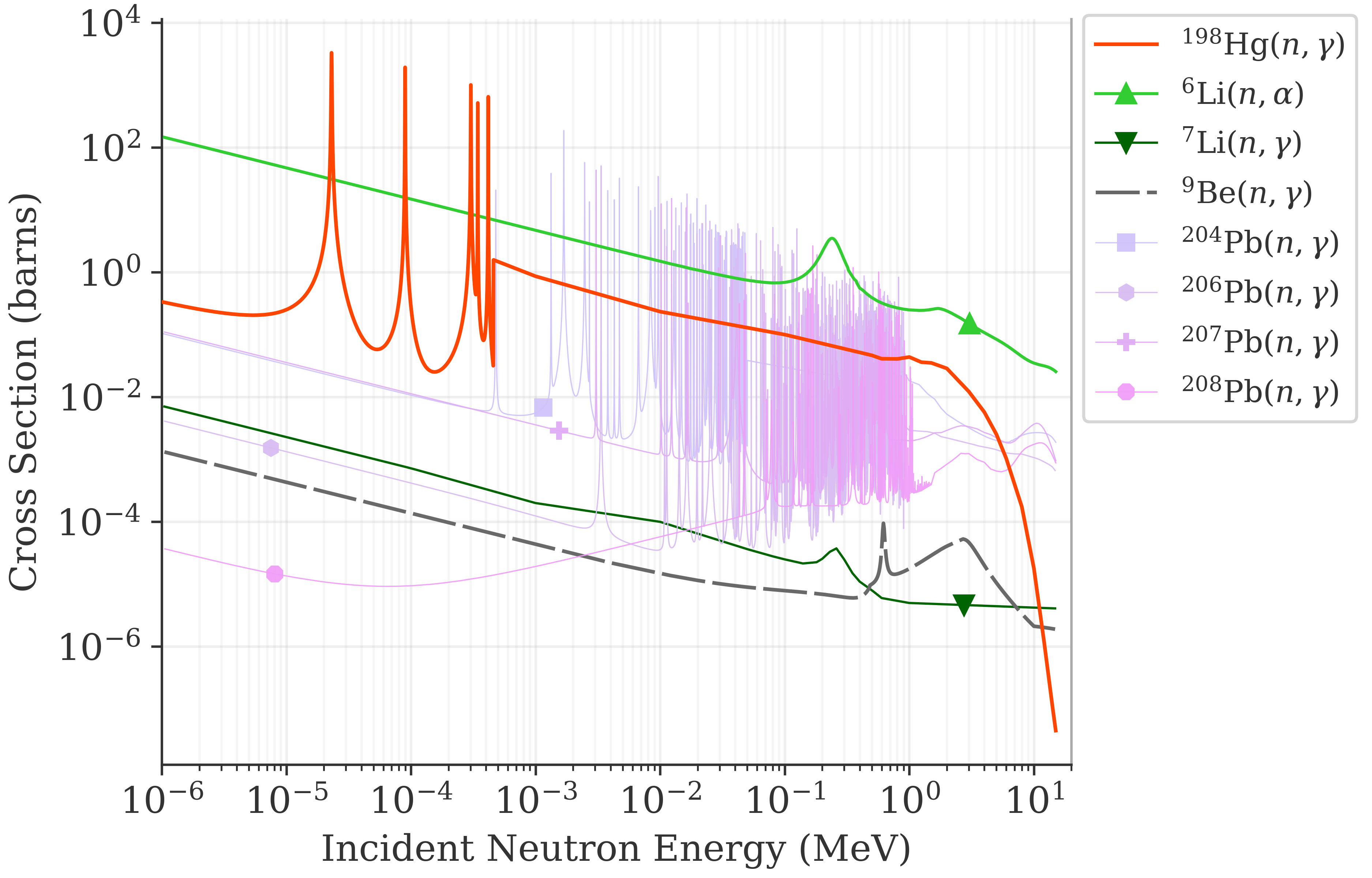}
  \caption{Cross sections for the $(n, \gamma)$ neutron capture cross sections in lead, \ce{^7Li}, \ce{^{198}Hg}, and \ce{^9Be}, as well as the cross section for the $\ce{^6Li}(n, \alpha)\ce{T}$ tritium-producing cross section.  The \ce{^{198}Hg} capture cross section is generally between that of lead and $\ce{^6Li}(n, \alpha)\ce{T}$, so most neutrons will still be captured by \ce{^6Li}. Data are from \cite{ENDFB8p1_Be9_ng,ENDFB8p1_Li7_ng,ENDFB8p1_Pb204_ng,ENDFB8p1_Pb206_ng,ENDFB8p1_Pb207_ng,ENDFB8p1_Pb208_ng}.}
  \label{fig:ngammaxsections}
\end{figure}

It is important that for the proposed feedstock material, the $(n, \gamma)$ neutron capture cross section is not so large that this reaction consumes too many thermal neutrons.
This reaction also creates gamma rays that can in some devices be one of the primary sources of heating in the superconducting magnets \cite{Hartwig2012}.
Figure~\ref{fig:ngammaxsections} shows the $(n, \gamma)$ cross sections for \ce{^{198}Hg} alongside the other neutron multipliers commonly proposed, as well as the tritium-producing $\ce{^6Li}(n,\alpha)\ce{T}$ reaction.  While the neutron capture cross section is substantially larger for \ce{^{198}Hg} than for any of the traditionally proposed neutron multiplier materials, we note that use of this material only in the inner blanket means that materials in the outer blanket layer can be selected to help shield magnets from the additional gamma rays. Additional shielding optimized for blocking gamma rays with high atomic number materials can also be used between the magnets and the blanket.
We also note that despite the relatively large neutron capture cross section for \ce{^{198}Hg}, capture of thermal neutrons in the system will still be dominated by the $\ce{^6Li}(n,\alpha)\ce{T}$ reaction, which has much larger cross section than all of the other reactions shown (with very local exceptions around resonances), as long as there is some meaningful fraction of \ce{^6Li} in the system.   

The inner blanket is also designed to flow the feedstock material out of the power plant core for removal of heat, tritium, and gold in downstream processing systems.  Since the proposed working fluid is a liquid metal, additional work will be needed to ensure that MHD pressure drops are acceptable and that any transient magnetic fields used for plasma control are not shielded from the plasma in confinement schemes where these are needed.  However, given the prevalence of Li and LiPb in blanket concepts, it is expected that these issues will be solvable.  

Existing methods can be used to getter the gold produced in the multiplier stream in a downstream processing loop using materials that more readily alloy with gold than mercury, including copper or tantalum \cite{Neuhausen2010Beijing}.  

The outer blanket in this configuration is similar to a traditional fusion blanket, and could be composed of any of the commonly used blanket working fluids (liquid Li, FLiBe, or LiPb).  Because neutrons in this outer portion of the device will be at lower energies, enrichment in \ce{^6Li} will likely benefit overall blanket design and assist in reducing the overall radial build for a target TBR.  

\subsection{OpenMC Monte Carlo Simulations}
\label{sec:openmc}

In this section, we first describe OpenMC simulations of TBR and gold production.  In Section \ref{sec:depletion} we will show the results of depletion simulations indicating what other species are generated, and in \ref{sec:activity} we will consider the required cooldown time for the gold produced from this process.  

To check the heuristic calculation of Section \ref{sec:multiplierrequirements} against a detailed neutronics code, we use  the neutronics code OpenMC \cite{romano2015openmc} to model a fusion power plant blanket geometry approximating that of the high power density ARC-class tokamak described in reference \cite{Frank2022ARC}.  A workflow comprised of OpenMC, Paramak \cite{10.12688/f1000research.28224.1}, openmc-plasma-source \cite{DelaporteMathurin2023_openmc_plasma_source}, and cad-to-dagmc \cite{shimwell_cad_to_dagmc_2024}  was used to set up the simulations \cite{harter_2025_15053541}.

\begin{figure}[htbp]
  \centering
  \begin{subfigure}[b]{0.48\textwidth}
    \centering
    \includegraphics[width=\textwidth]{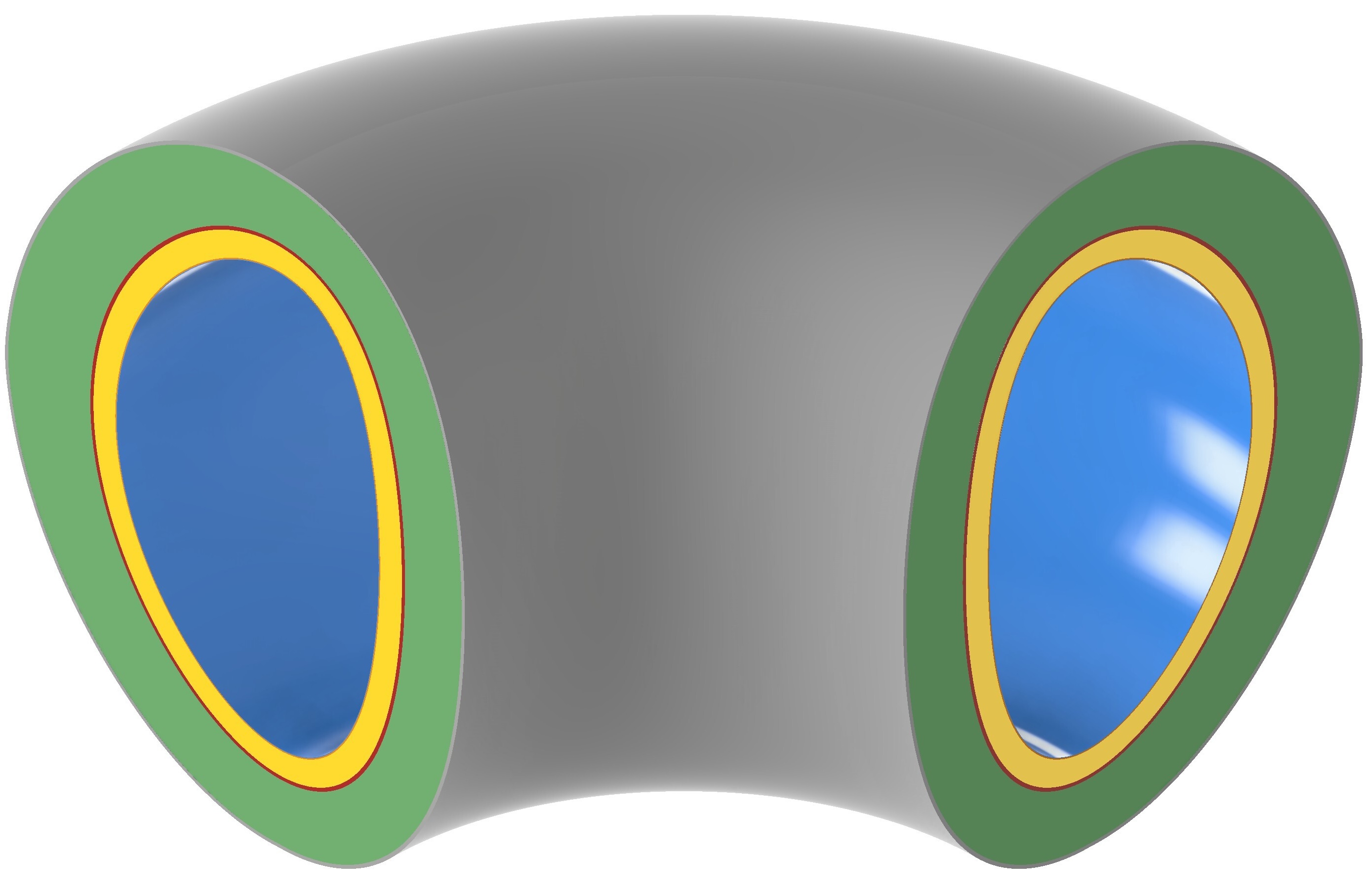}
    \label{fig:tokamak_left}
  \end{subfigure}
  \hfill
  \begin{subfigure}[b]{0.48\textwidth}
    \centering
    \includegraphics[width=\textwidth]{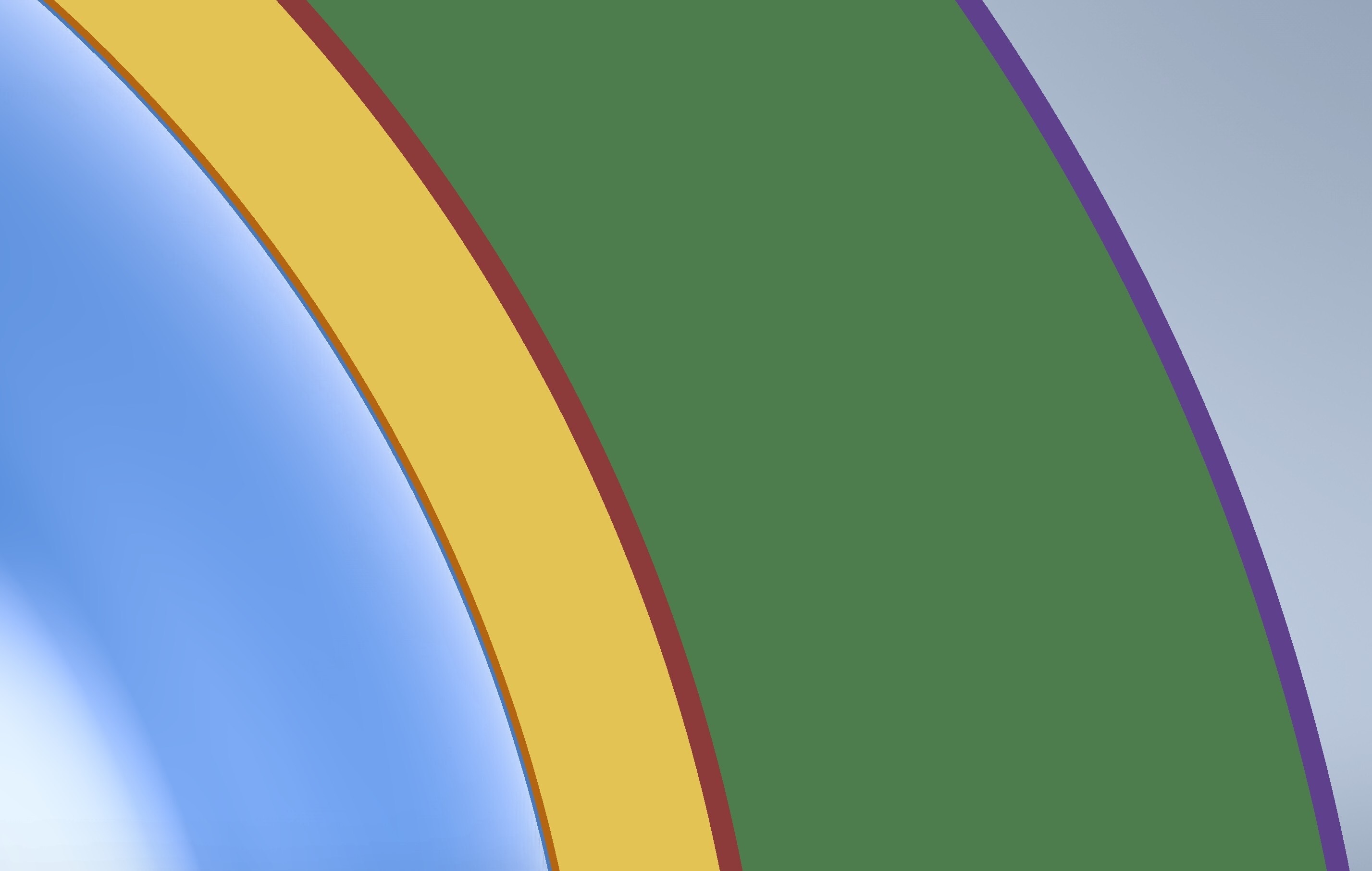} 

    \label{fig:tokamak_right}
  \end{subfigure}

  \caption{At left: the simplified geometry used for Monte Carlo neutronics simulations on an example tokamak blanket configuration implementing \ce{^{198}Hg} as a neutron multiplier. On the right is a zoomed view of the radial build, showing the armor and first wall structure (brown), \ce{Hg} multiplier layer (yellow), inner blanket structure (red), and the outer FLiBe blanket (green).}
  \label{fig:tokamak_pair}
\end{figure}

Paramak was used to create a set of tokamak geometries with a constant outer blanket diameter but varying channel and blanket thicknesses.  The first wall was chosen to be \SI{5}{mm} of Tungsten, \SI{10}{mm} of V-4Cr-4Ti was used for the inner structural layer, \SI{30}{mm} of V-4Cr-4Ti for the outer structural layer, and the blanket tank was \SI{30}{mm} of Eurofer97 \cite{STORNELLI20245075} steel.  For runs shown here, the power plant is assumed to have $1500\,\mathrm{MW_{th}}$ power output, following the design point in \cite{Frank2022ARC}. The tool cad-to-dagmc was used to export these geometries for use in OpenMC.

The inner channel thickness was scanned from \SI{5}{mm} up to \SI{350}{mm} in steps of \SI{5}{mm} between \SI{5}{mm} and \SI{150}{mm} and then in steps of \SI{10}{mm} between \SI{150}{mm} and \SI{350}{mm}. The outer blanket thickness changed in proportion to the channel with the inboard side blanket thickness ranging from \SI{750}{mm} down to \SI{405}{mm} and the outboard thickness ranging from \SI{1000}{mm} down to \SI{655}{mm}. The geometric parameters of the simulation are summarized in Table \ref{tab:tokamak-parameters}. Neutronics simulations were performed using these geometries for varying mixtures of an enriched $\ce{LiHg}$ channel material. The Hg concentrations simulated ranged from $0-100\mathrm{at}\%$ in $5\mathrm{at}\%$ increments with $\ce{Li}$ composing the balance of the material. For simplicity, in this model we assume that the two materials are mixed, but note that in a practical blanket design these fluids do not need to be in direct contact, and structural material can be used to separate the two fluids while preserving similar neutronics properties to the system modeled here. 
All Hg was isotopically enriched to $90\mathrm{at}\%$ $\ce{^{198}Hg}$ and all $\ce{Li}$ was enriched to $90\mathrm{at}\%$ $\ce{^6Li}$. The remaining $10\mathrm{at}\%$ of $\mathrm{Hg}$ was for simplicity assumed to be composed of the natural abundance mixture of isotopes except $\ce{^{198}Hg}$; depending on the specific enrichment approach used the remaining isotopes might have different abundances to those used here.  Modeling of isotopic mixtures more representative of specific enrichment schemes is left as future work.  The channel temperature was assumed conservatively to be 900K, and density of the LiHg mixture was calculated based on the independent densities of Li and Hg, neglecting mixing effects for simplicity.  

\begin{table}[ht]
    \centering
    \begin{tabular}{@{}l r@{}}
        \toprule
        \textbf{Parameter} & \textbf{Value} \\ \midrule
        Major radius $R$ & 4.2\,m \\
        Minor radius $a$ & 1.2\,m \\
        Plasma elongation $\kappa$ & 1.6 \\
        Plasma triangularity $\delta$ & 0.25 \\
        First-wall thickness & 5\,mm \\
        Inner structural layer & 10\,mm \\
        Channel thickness & 5–350\,mm \\
        Outer structural layer & 30\,mm \\
        Outboard blanket thickness & 655–1000\,mm \\
        Inboard blanket thickness & 405–750\,mm \\
        Blanket tank thickness & 30\,mm \\ \bottomrule
    \end{tabular}
    \caption{Key tokamak parameters.}
    \label{tab:tokamak-parameters}
\end{table}

The neutron source used was a 'FusionRingSource' object generated using the package OpenMC-plasma-source. All simulations were run using a 90 degree segment of a tokamak with reflective boundaries at 0 and 90 degrees to decrease computational intensity while maintaining accuracy. The nuclear data library used for this OpenMC setup was JENDL-5 \cite{Iwamoto02012023} due to its higher level of detail for certain \ce{Au} and \ce{Hg} isotopes of interest. These files were prepared for use in the simulation using OpenMC’s nuclear data API operating with NJOY2016 \cite{osti_1338791}. 

\begin{figure}[htbp]
  \centering
  \includegraphics[width=0.9\textwidth]{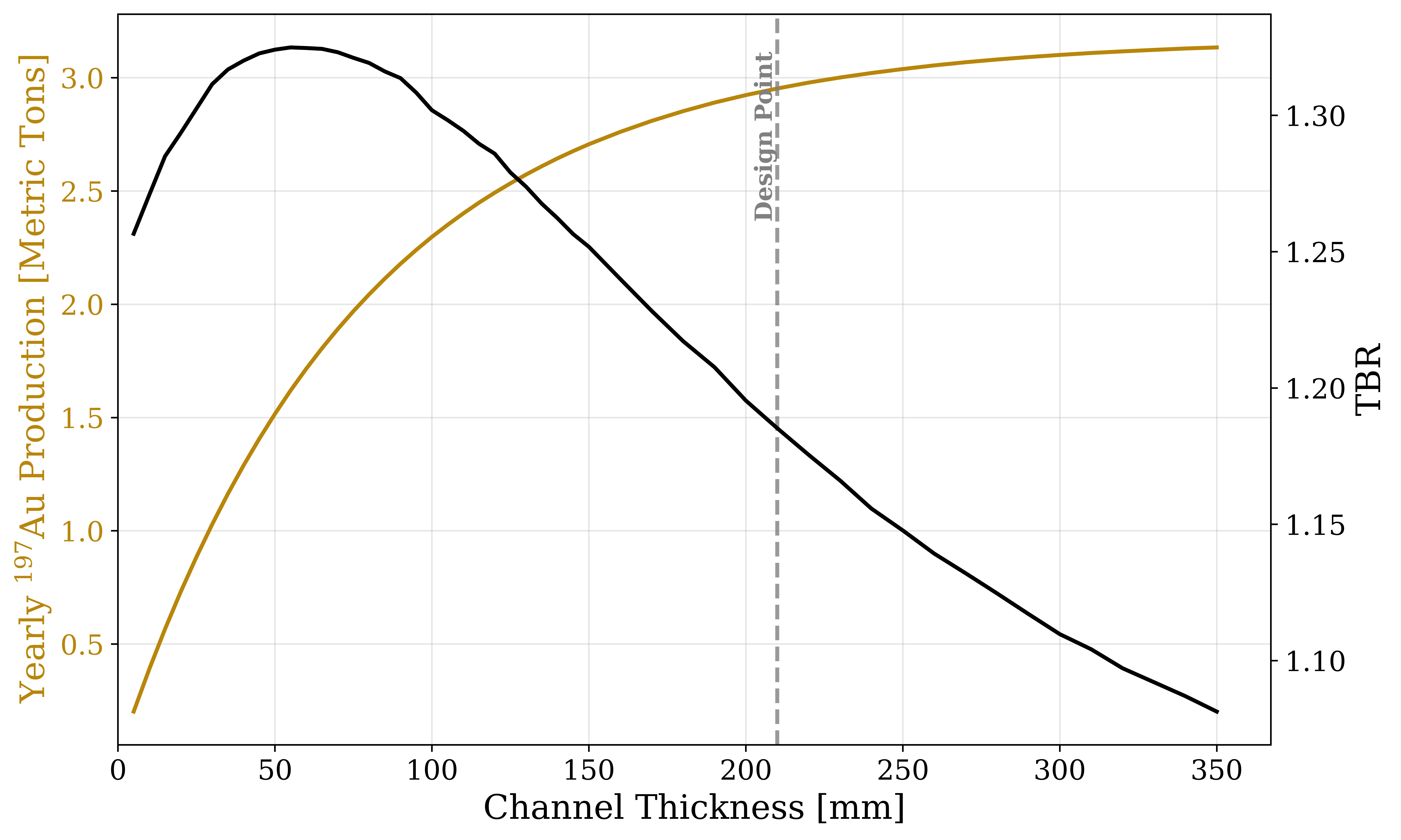}
  \caption{Tritium breeding ratio (TBR) and yearly stable gold production as a function of multiplier channel thickness for a mixture of $85\mathrm{at}\%$ $\ce{Hg}$ and $15\mathrm{at}\%$ $\ce{Li}$, where each is assumed to be $90\%$ enriched in the target isotope.}
  \label{fig:1dplot_au_tbr}
\end{figure}

A simple one-dimensional scan on multiplier layer thickness can be seen in Figure~\ref{fig:1dplot_au_tbr} for a fixed choice of $85\mathrm{at}\%$ \ce{Hg} and $15\mathrm{at}\%$ \ce{Li}, illustrating some of the basic tradeoffs at this particular multiplier composition.  Because the mercury multiplier layer provides additional neutrons, initial increases in this layer thickness increase the TBR.  Eventually, as most of the possible $(n,2n)$ reactions are achieved, further increases in the multiplier thickness do not provide substantial additional neutrons and the $\ce{^6Li}$ enriched FLiBe layer is more effective than the multiplier layer at producing tritium from the thermal neutrons, so further increases in multiplier layer thickness result in decreasing TBR with the gold production saturating. 

\begin{figure}[htbp] 
  \centering
  \includegraphics[width=0.9\textwidth]{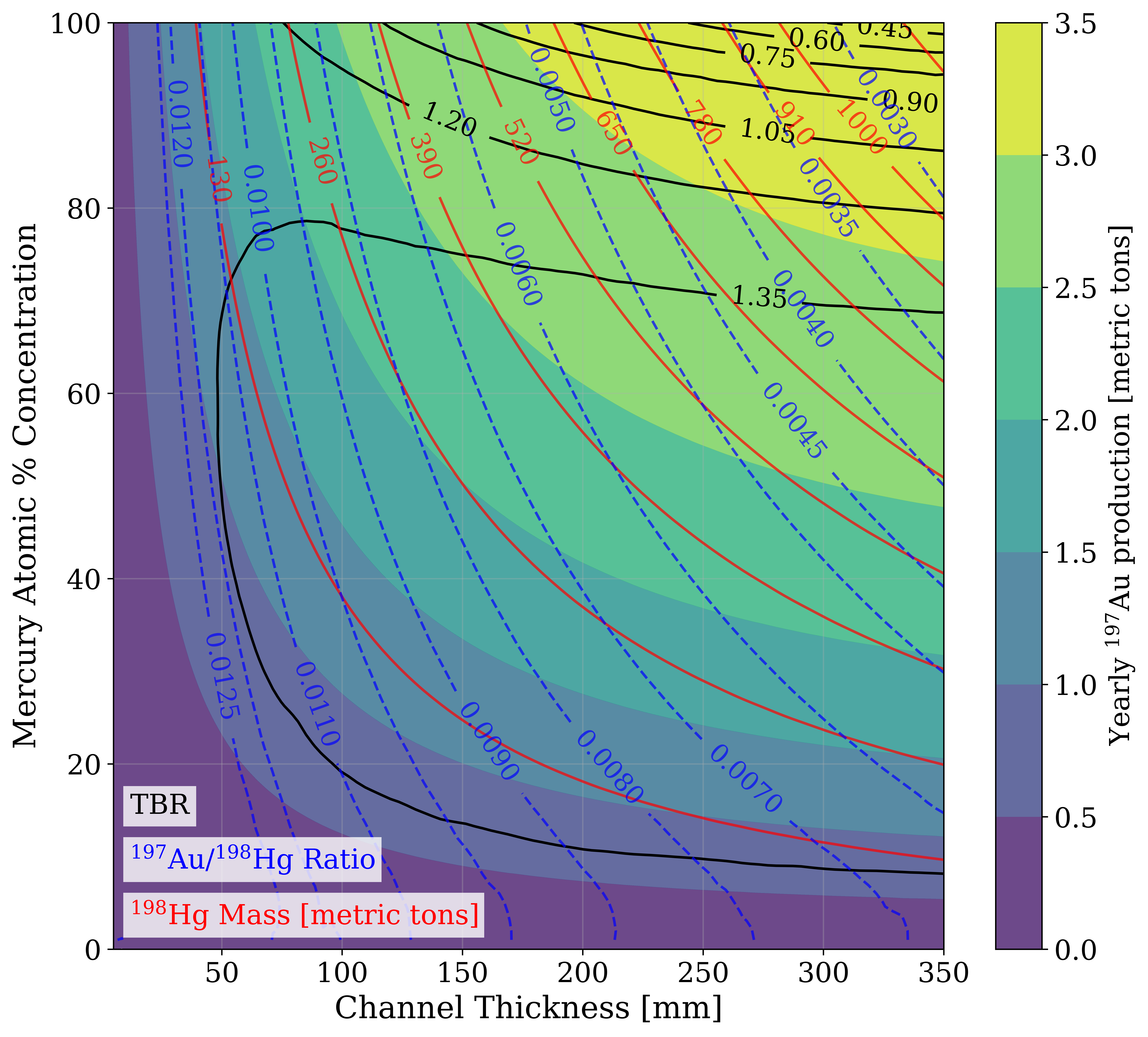}
  \caption{Blanket TBR (black contours), \ce{^{197}Au} production rate (color scale), initial $\ce{^{198}Hg}$ loading (red contours), and the ratio of yearly \ce{^{197}Au} production to the initial $\ce{^{198}Hg}$ loading (dotted blue contours) in the simplified power plant configuration shown in Figure~\ref{fig:tokamak_pair}.}
  \label{fig:blanketscan}
\end{figure}

To more fully assess the impacts of $\ce{Li}$ fraction in the multiplier layer, we also plot an additional scan parameter along with more evaluation metrics. Figure~\ref{fig:blanketscan} maps out the yearly $\ce{^{197}Au}$ production alongside the TBR, initial $\ce{^{198}Hg}$ inventory in the multiplier layer, and the yearly ``burn efficiency'' of the layer, defined as the mass of $\ce{^{197}Au}$ produced in a year divided by the mass of initial $\ce{^{198}Hg}$ inventory.  

Multiple configurations are capable of producing $>\SI{3000}{kg/yr}$ (\SI{2000}{kg/GW_{th}/yr}) of \ce{^{197}Au} while maintaining a $\mathrm{TBR}>1$, meeting fuel cycle requirements while still producing large quantities of gold. For many of the cases simulated the value of the \ce{^{197}Au} produced was larger than the value of all electricity produced by the power plant over the same period. 

We note now another trade-off: the highest TBR values are achieved for thick multiplier layers with higher \ce{Li} fraction, but this choice results in lower \ce{^197Au} production. Yet another tradeoff exists between \ce{^{197}Au} production and the initial inventory of \ce{^{198}Hg}.  Because the average neutron energy decreases exponentially with distance into the blanket, thinner layers of mercury are more efficient at gold production when normalizing to the initial loading.  A reasonable range of initial mercury loadings that still achieves high rates of gold production is 100--450\,\si{t/GW_{th}}.  

The simulated case of a \SI{210}{mm} thick channel with an $85 \mathrm{at}\%$ $\ce{Hg}$ concentration was identified as one of the more desirable potential tokamak setups based on consideration of the above tradeoffs, and was selected as a specific reference case for depletion simulations described in Section \ref{sec:depletion}. This configuration produced \SI{2953}{kg/yr} of \ce{^{197}Au} while maintaining a total TBR of 1.19 of which 0.490 resulted from the FLiBe blanket and 0.694 from the \ce{LiHg} channel. The other $\approx0.006$ is from other layers within the tokamak.
The mass ratio of \ce{^{197}Au} produced over one year to the initial inventory of \ce{^{198}Hg} was 0.0047, with an overall \ce{^{198}Hg} mass of 622 tons.

This work aims to perform an analysis for a representative geometry, with rough estimates for the yearly \ce{^{197}Au} generation rate.  We note that there are many design optimizations that can be used to improve overall performance of the design.  Use of armor and structural materials that are more transparent to neutrons could improve product generation.  Implementing an additional beryllium multiplier layer just outside the \ce{^{198}Hg} multiplier layer could provide additional multiplication for TBR purposes from the lower energy neutrons due to its lower threshold for $(n,2n)$ reactions. Spin-polarized fuel \cite{kulsrud1986spin} could also be used to achieve neutron emission preferentially aligned with the magnetic field orientation, easing the blanket design challenge \cite{Borowiec2024SPF}.  While this work studies a fixed outer blanket outer radius, we also note that further optimizations could investigate the potential to decrease the radial build thickness of the device while maintaining target TBR and gold production rates. 

Ultimately, design of blankets optimized for \ce{^{197}Au} generation will be done with respect to the overall system economics, balancing consideration for gold production, blanket thickness, TBR, \ce{^{198}Hg} inventory, and magnet heating.  While a challenging problem, it represents only a moderate increase in complexity on top of the existing blanket design problem, and is well justified by the additional economic benefit.  

\subsubsection{Blanket Depletion Simulations}
\label{sec:depletion}

\begin{table}[b]
  \centering

\begin{subtable}[b]{0.48\textwidth}
    \centering
    \begin{tabular*}{\linewidth}{@{\extracolsep{\fill}} l r r}
      \toprule
      Isotope & \multicolumn{1}{c}{Mass (g)} & Half-Life (yr) \\
      \midrule
      $^{197}\mathrm{Au}$ & 2.91E+06 & Stable \\
      $^{198}\mathrm{Hg}$ & 2.10E+03 & Stable \\
      $^{195}\mathrm{Au}$ & 2.89E+02 & 5.10E-01 \\
      $^{195}\mathrm{Pt}$ & 2.25E+02 & Stable \\
      $^{196}\mathrm{Pt}$ & 1.64E+02 & Stable \\
      $^{198}\mathrm{Au}$ & 9.34E+01 & 7.37E-03 \\
      $^{199}\mathrm{Hg}$ & 4.52E+01 & Stable \\
      $^{196}\mathrm{Hg}$ & 1.24E+01 & Stable \\
      $^{196}\mathrm{Au}$ & 9.60E+00 & 1.69E-02 \\
      $^{199}\mathrm{Au}$ & 2.21E+00 & 8.60E-03 \\
      \bottomrule
    \end{tabular*}
  \end{subtable}
  \hfill
  \begin{subtable}[t]{0.48\textwidth}
    \centering
    \begin{tabular*}{\linewidth}{@{\extracolsep{\fill}} l r r}
      \toprule
      Nuclide & \multicolumn{1}{c}{Mass (g)} & Half-Life (yr) \\
      \midrule
      $^{198}\mathrm{Hg}$ & 6.19E+08 & Stable \\
      $^{6}\mathrm{Li}$   & 3.22E+06 & Stable \\
      $^{202}\mathrm{Hg}$ & 2.34E+07 & Stable \\
      $^{200}\mathrm{Hg}$ & 1.80E+07 & Stable \\
      $^{199}\mathrm{Hg}$ & 1.56E+07 & Stable \\
      $^{7}\mathrm{Li}$   & 4.32E+05 & Stable \\
      $^{201}\mathrm{Hg}$ & 1.04E+07 & Stable \\
      $^{204}\mathrm{Hg}$ & 5.43E+06 & Stable \\
      $^{4}\mathrm{He}$   & 7.73E+04 & Stable \\
      $^{3}\mathrm{H}$    & 5.64E+04 & 1.23E+01 \\
      $^{196}\mathrm{Hg}$ & 1.14E+05 & Stable \\
      $^{3}\mathrm{He}$   & 1.57E+03 & Stable \\
      $^{203}\mathrm{Tl}$ & 5.54E+04 & Stable \\
      $^{1}\mathrm{H}$    & 1.67E+02 & Stable \\
      $^{197}\mathrm{Hg}$ & 3.10E+04 & 7.33E-03 \\
      $^{203}\mathrm{Hg}$ & 1.25E+04 & 1.28E-01 \\
      $^{205}\mathrm{Tl}$ & 5.33E+03 & Stable \\
      $^{195}\mathrm{Pt}$ & 5.07E+02 & Stable \\
      $^{2}\mathrm{H}$    & 2.24E+00 & Stable \\
      $^{196}\mathrm{Pt}$ & 1.32E+02 & Stable \\
      $^{204}\mathrm{Tl}$ & 9.95E+01 & 3.77E+00 \\
      $^{194}\mathrm{Pt}$ & 2.49E+01 & Stable \\
      $^{202}\mathrm{Tl}$ & 1.24E+01 & 3.36E-02 \\
      $^{198}\mathrm{Pt}$ & 5.54E+00 & Stable \\
      $^{204}\mathrm{Pb}$ & 5.32E+00 & 1.39E+17 \\
      $^{206}\mathrm{Pb}$ & 3.16E+00 & Stable \\
      \bottomrule
    \end{tabular*}
\end{subtable}
  \caption{Gold isotopes removed and resultant decay products (left) and final channel nuclide inventory (right) resulting from a one year depletion run of a \SI{1500}{MW} power plant with a \SI{210}{mm} channel and $85\mathrm{at}\%$ $\ce{Hg}$.  We list only isotopes with $>\SI{1}{\gram}$ of material remaining. The nuclides shown are specifically those in the multiplier channel, and do not list any materials generated in the FLiBe ``outer blanket''.}
  \label{tab:nuclide_inventory}
\end{table}

The initial OpenMC simulations provided a detailed model of the \ce{^{197}Au} production rate from a set of incident neutron reactions across different Hg isotopes. From this it was confirmed that $\ce{^{198}Hg}$ $(n,2n)$ is by far the dominant reaction for \ce{^{197}Au} production. A full list of the \ce{^{197}Au} producing reactions simulated is available in Appendix~\ref{sec:Auxsections}. Depletion simulations characterize other isotopes produced in the system, show the effects of longer chain pathways on total $\ce{^{197}Au}$ production, and quantify production rates of the longest-lived mercury isotope $\ce{^{194}Hg}$ and the longest-lived unstable gold isotope $\ce{^{195}Au}$.

Coupled particle transport and depletion simulations were performed for the case of \SI{210}{mm} channel thickness with $85\mathrm{at}\%$ mercury concentration as described in Section~\ref{sec:openmc}. These were done using OpenMC’s CoupledOperator with the CF4 integrator, the neutron source object was the same as the initial simulations, and the power plant power was \SI{1500}{MW} as before. The depletion chain file used for these simulations was generated from the JENDL-5 data library using the OpenMC API. To model a full year of continuous operation, 24 timesteps of 15.22 days were used. In order to approximate a continuous extraction of the Au produced in the system, between each depletion timestep all $\ce{Au}$ atoms present were removed from the channel material and replaced with an identical amount of $90\mathrm{at}\%$ enriched Hg.

After a simulated year the total \ce{^197Au} removed was \SI{2909}{kg}, close to the pure particle transport value of \SI{2953}{kg} for the same setup. The total mass of $\ce{^{195}Au}$ removed was \SI{289}{g}. The \ce{^194Hg} production is deemed negligible with only around \SI{4.31}{mg} being produced in the year-long cycle. Table  \ref{tab:nuclide_inventory} shows the mass of $\ce{Au}$ isotopes removed and the final composition of the channel material; for these values, decay after removal was considered, explaining the difference in amount with the stated removal values for $\ce{^{195}Au}$. 

We expect that the long-lived waste challenge for this material will be driven by the presence of $\ce{^{204}Tl}$, the longest-lived isotope of thallium with a half life of 3.78~years~\cite{kaeri_tl204}, but only \SI{99.49}{g} of this material were produced in the configuration shown.  Isotopic enrichment of the feedstock mercury can be used to decrease the quantity produced if deemed desirable by the power plant designer.  

\subsubsection{Activity of Gold Product}
\label{sec:activity}

Any residual radioactivity of the gold produces through this process has the potential to affect the value of the product.  Fortunately, the longest-lived isotope of gold is relatively short lived with a half-life of $186.10$ days \cite{kaeri_au195}, meaning that a fairly short cooling time will be required to introduce this gold to the market. In addition, this isotope is generated through $(n,2n)$ reactions on the already low abundance $\ce{^{196}Hg}$, producing $\ce{^{195}Hg}$ that decays to $\ce{^{195}Au}$.  As a result the cooling time can be reduced by isotopically purifying the feedstock to remove $\ce{^{196}Hg}$.  Still, for any practical amount of isotope separation some cooldown period will still be required.

To be categorized as Class-A low level waste (the least hazardous classification), the Nuclear Regulatory Commission (NRC) requires that material have $<700\ \mathrm{Ci/m^{3}}$ activity for all nuclides with less than a five year half-life \cite{USNRC_61_55}.  For the mixture of gold isotopes identified in Table \ref{tab:nuclide_inventory}, it takes about $6.8$ years for the activity concentration to fall below this level. 

To not require any form of labeling as radioactive, the product gold has to have less radioactivity than the threshold for Class 7 waste. NRC rules \cite{CFR173436} consider $\ce{^{197}Au}$ to be Class 7 when activity concentration is $>\SI{2700}{pCi/g}$, which is reached after 13.7~years for the initial concentration listed.  

An even more stringent constraint can be applied for any gold that will be regularly handled by the general population.  As a highly conservative requirement, we can stipulate that this gold must be less radioactive than a banana.  Due to \ce{^{40}K} content, bananas have an activity of $\sim 3520\,\mathrm{pCi/kg}$, or about $420 \,\mathrm{pCi}$ for a single banana. To meet this requirement, a troy ounce of gold with the initial isotope mix shown in Table \ref{tab:nuclide_inventory} must sit for about 17.7~years to be below a banana equivalent level of activity.  

In practice, given that much of all gold is used to store value and is not actively in use, we do not expect the need to store it for 7--17 years to be a major impediment; at worst, it means that the product will initially have somewhat less value than pure $\ce{^{197}Au}$, and so some discount should be applied to the value of freshly produced gold.  However, we also note that our assumption around initial $\ce{^{196}Hg}$ inventory used here is conservative, and for every factor of ten decrease in starting $\ce{^{195}Au}$ quantity, about 1.7 years is saved on cooldown time for the values given below.  As such, the optimal cooldown time can ultimately be viewed as an economic tradeoff between the cost of additional isotopic purification of feedstock mercury and the cost of delaying gold transport or sale. 
 
\section{Discussion} \label{sec4}

In this section we compare mercury with other multiplier materials, address some of the key challenges on the path to deployment of this approach, and discuss the implications of this work for the economic viability of fusion. 

\subsection{Comparing Mercury with Other Multipliers}

We first note that while use of mercury as a neutron multiplier introduces some unique challenges, it also mitigates some of the key concerns with traditionally chosen neutron multiplier candidates. In Table \ref{tab:blanket-materials} we provide a qualitative comparison of the traditionally selected multiplier materials alongside mercury.

\newcommand{\Good}{\textcolor{green!60!black}{\ding{51}}} 
\newcommand{\Bad}{\textcolor{red}{\ding{55}}}             
\newcommand{\Maybe}{%
  \textcolor{orange!85!black}{%
    \raisebox{-0.85ex}{\smash{\small\textasciitilde}}%
  }%
}

\begin{table}[htbp]
  \centering
  \caption{Qualitative comparison of candidate neutron-multiplier/blanket materials. A green check indicates good properties, a red "x" indicates bad properties, and the orange "$\sim$" indicates acceptable but non-ideal performance. }
  \label{tab:blanket-materials}
  \renewcommand{\arraystretch}{1.2} 
  \begin{tabular}{@{} l c c c c @{}}
    \toprule
    & \textsuperscript{7}Li & Be/FLiBe & Pb & Hg \\
    \midrule
    $(n,2n)$ or $(n,n{+}T)$ cross section        & \Good & \Good & \Good & \Good \\
    Small $(n,\gamma)$ cross section             & \Good & \Good & \Maybe & \Maybe \\
    Abundance,  cost                           & \Good & \Bad  & \Good & \Good \\
    Activation products                          & \Good & \Good & \Bad  & \Good \\
    Reactivity/compatibility                                   & \Bad  & \Maybe & \Good & \Good \\
    Toxicity (before or after irradiation) & \Good & \Bad  & \Bad  & \Maybe \\
    \bottomrule
  \end{tabular}
\end{table}

As already discussed, the key neutronics properties of mercury are comparable with those of other common multiplier options, albeit with a higher but still acceptable neutron capture cross section. For other engineering requirements, mercury actually has superior performance to any of the other common multiplier materials, with higher abundance than beryllium, a smaller quantity of long lived activation products than lead, and lower chemical reactivity than lithium. Lastly, while mercury poses its own safety challenges, elemental mercury should be significantly easier to handle than beryllium---the permissible exposure limit for beryllium is only \SI{0.2}{\micro\gram/m^3} over an 8 hour time-weighted average \cite{OSHA2021_BerylliumFactSheet}, while the permissible exposure level for mercury vapor is \SI{25}{\micro\gram/m^3} over the same period \cite{OSHA2022_MercuryVapor}, a difference of two orders of magnitude. Safety challenges associated with mercury are discussed briefly in Section~\ref{sec:regsafety}.

%
%
%

We also note that while use of mercury as a neutron multiplier leads to different engineering constraints than more conventional materials, it is not necessarily worse than other materials.  As one example, mercury has a high vapor pressure relative to other materials.  While this could constrain the operating temperature of the multiplier layer, the high vapor pressure could also be used as a tool to drive a turbine \cite{Gutstein_Furman_Kaplan_1975} or an MHD generator \cite{Hoffman_Campbell_Logan_1988}, allowing for a form of unconventional power conversion. Another possibility is to leverage the high vapor pressure of mercury by using evaporative cooling to reduce the required MHD pumping power in the system. Another perspective is techno-economic in nature: when fusion systems make a co-product equal in value to the electricity output of the system, it is no longer necessary to operate blankets at extreme temperatures to increase power conversion efficiency.  By relaxing these engineering requirements, early deployments of fusion systems could potentially be accelerated.
Future work will address integrated system design questions to demonstrate blanket systems engineered around mercury use.  

\subsection{Implementation Challenges}

The two most important technical challenges to implementing the approach described here are scaling mercury isotope separation, and demonstrating blanket structural materials compatible with both $\ce{Li}$ and $\ce{Hg}$ at the required operating conditions for a fusion device.  Deployment of this technology will also need to ensure safe usage of mercury and conform to the regulatory constraints controlling this material.  Both the isotope separation challenge and the regulatory challenge will be strongly driven by the inventory of mercury required for each system, so we begin with a consideration of this topic.  

\subsubsection{Mercury Inventory Reduction}
\label{inventorysubsection}

It is important to note that the initial $\ce{^{198}Hg}$ loading of the tokamak power plants described here is fairly large (100--400 tons/\si{GW_{th}}), and \ce{^{198}Hg} feedstock is converted to \ce{^{197}Au} product at a relatively low rate, $\sim 0.3$\textendash $1.2\%/\mathrm{yr}$ depending on the specific design point chosen, as seen in Figure~\ref{fig:blanketscan}. We note that because of the decrease of neutron energy as a function of distance into the blanket, the initial loading required for a given transmutation efficiency can be heuristically estimated to scale like a fixed wall thickness multiplied by the surface area to be covered to collect all desired neutrons.  This scaling favors fusion concepts with high volumetric power density; inertial and magnetoinertial concepts may be the best positioned to benefit from this approach if $\ce{^{198}Hg}$ feedstock supply is limited or costly.  

A simple limiting case can be considered to estimate a lower bound on multiplier inventory which could achieve similar gold generation rates to the design point considered here.  For a highly compact fusion source such as in ICF (effectively a point source for our purposes), the multiplier volume is, in the limit, a spherical shell around this point with radius of \SI{210}{mm}.  This corresponds to a volume of $39 \ \mathrm{liters}$, or a mercury mass of about \SI{525}{kg}. In practice, a meaningful mass of mercury will still be needed in the mercury processing and heat extraction systems, but including some conservatism it is not unrealistic to expect that full systems could use less than $6\,\mathrm{tons}$ of~\ce{^{198}Hg}, two orders of magnitude lower than the tokamak design point identified above.  


While there is no clear way that magnetic confinement concepts could reach inventories this low, other tools can be used to increase the neutron wall loading in certain regions of some fusion concepts.  By increasing localized wall loading, thicker regions of mercury feedstock can be used in the regions where these layers will have the most benefit, and overall inventories could be reduced.  As previously mentioned,  use of spin polarized fuel enables directional emission of neutrons and can locally increase neutron wall loading.  

Magnetic mirrors can also be imagined as a specific magnetic confinement concept capable of achieving higher local neutron wall loading to achieve more efficient mercury feedstock usage.  Specifically, because the fuel density profile is highly peaked at the sloshing ion bounce points, the neutron emission will also be localized in these regions and higher mercury burn efficiency could in principle be achieved there. As a specific example, studies of a neutron source based on the gas dynamic trap (GDT) simulated neutron wall loading peaking by about a factor of six at the sloshing ion turning points above the value at the midplane \cite{Fischer2000, Anikeev2015}. 

Because of geometric complexity of their vacuum vessel, stellarators can experience highly nonuniform neutron wall loading varying by more than an order of magnitude between different locations \cite{Najmabadi2008}.  While typically considered a downside of this approach, this spatial nonuniformity could be used as a beneficial tool in this case to more efficiently use thicker mercury feedstock layers in the high flux regions while potentially reducing the overall feedstock mass required.  

\subsubsection{Mercury Isotope Separation}

Mercury isotope separation has already been demonstrated with a number of different technical approaches.  Early work investigated a photochemical approach based on the use of isotopically enriched $\mathrm{Hg}$ light sources to selectively excite certain isotopes and preferentially remove them through chemical reactions \cite{webster_photochemical_nodate, Billings1953Photochemical}. Reference \cite{Vyazovetskii1998HgIsotopes} specifically identifies a single-step photochemical method to enrich \ce{^{198}Hg}, and demonstrates enrichment up to 99.2\% in only four separation stages.  

Another more recent approach has proposed the use of selective photoexcitation combined with deflection by magnetic fields to achieve the desired separation \cite{raizen_mercury_2016}.  

To gauge the cost of these photon-driven separation approaches, we estimate 10 photons per excitation provided in \cite{raizen2012magnetically} and supported by the range of photonic efficiencies of 0.026--0.66 given in \cite{Noel2017_CFR}, $33\%$ wall plug efficiency for mercury vapor lamps based on \cite{fredes2021estimation}, and $85\%$ of the emitted light goes into photons at \SI{254}{nm}, as described by \cite{Kowalski2009_UVGILamps}.
To selectively excite only the product material requires $\sim\SI{270}{kW}$ of wall-plug power to process 100 tons of feedstock \ce{^{198}Hg} in a year, corresponding to $\sim \$\SI{2.4}{\per\kilo\gram}$ of \ce{^{198}Hg} product at the conservative assumption of $\$0.1\mathrm{/kWh}$ (in practice, this process could be done behind the grid in a fusion plant for lower cost per kWh).  Multiple stages may be desired for this process depending on the enrichment target, requiring excitation of more material and higher associated energy costs; however, the estimate included here provides a reasonable starting point with conservative assumptions included in both the photon count and energy cost.  Aside from operational expenditures (OPEX) due to energy usage, capital expenditures may be roughly estimated based on the price per watt of commercially available mercury sterilization lamps. Microwave-driven lamps similar to those used in prior work are available for $\sim \$0.06\mathrm{/W} $\cite{cureuv558492}, with microwave power sources available for $\sim \$0.065\mathrm{/W}$ \cite{fengyu_microwave_2kw245ghz} for scientific supplies, or much less for commercial microwave ovens ($\sim \$0.015\mathrm{/W} $ \cite{alibaba_magnetron_1000w}). Even assuming a conservative value of $\$1\mathrm{/W}$ for the driver, the capital expenditures (CAPEX) represents $< \$1 \mathrm{M}$. Future work will examine isotope separation systems in more detail, but this brief estimate indicates that the costs for isotope separation appear robust to orders of magnitude of margin on OPEX and CAPEX.  

Early work on AVLIS at LLNL also used $\mathrm{Hg}$ as a feedstock material, demonstrating feasibility of this pathway as well \cite{Crane1986UCRL94164}.  More recent work has also demonstrated the use of centrifuges for separation, offering yet another option \cite{babaev_centrifugal_2010}. 

While details of scalable mercury isotope separation need to be fully defined, this represents a challenge of similar or lower magnitude to enrichment of \ce{^{6}Li} for fusion blankets.  The \SI{708}{MW_{th}} ARC \cite{sorbom_arc_2015} uses about 50 tons of \ce{^{6}Li} at in $90\%$ enrichment FLiBe in its blanket, or about $1.8\times 10^7 \ \mathrm{mol}$ for $1500\mathrm{MW_{th}}$, while the design point identified here (loosely optimized to balance gold production rate and mercury inventory) uses about 600 tons, or about $3\times 10^6 \ \mathrm{mol}$ for a \SI{1500}{MW_{th}} device.

\begin{table}[htbp]
  \centering
  \caption{Isotopes of mercury (\ce{_80Hg}) \cite{haynes2016crc}}
  \label{tab:HgIsotopes}
  \begin{tabular}{@{}lcccl@{}}
    \toprule
      & \multicolumn{2}{c}{\textbf{Main isotopes}} 
      & \multicolumn{2}{c}{\textbf{Decay}}\\
    \cmidrule(r){2-3}\cmidrule(l){4-5}
    \textbf{Isotope} & \textbf{Abundance} & \textbf{Half-life $(t_{1/2})$} & \textbf{Mode} & \textbf{Product}\\
    \midrule
    $^{194}$Hg & synth  & 444 y      & $\varepsilon$ & $^{194}$Au\\
    $^{195}$Hg & synth  & 9.9 h      & $\beta^{+}$   & $^{195}$Au\\
    $^{196}$Hg & 0.15\% & stable     & —            & —\\
    $^{197}$Hg & synth  & 64.14 h    & $\varepsilon$ & $^{197}$Au\\
    $^{198}$Hg & 10.0\% & stable     & —            & —\\
    $^{199}$Hg & 16.9\% & stable     & —            & —\\
    $^{200}$Hg & 23.1\% & stable     & —            & —\\
    $^{201}$Hg & 13.2\% & stable     & —            & —\\
    $^{202}$Hg & 29.7\% & stable     & —            & —\\
    $^{203}$Hg & synth  & 46.612 d   & $\beta^{-}$   & $^{203}$Tl\\
    $^{204}$Hg & 6.82\% & stable     & —            & —\\
    \bottomrule
  \end{tabular}
\end{table}

Not only does this separation require far fewer moles of material than $\ce{^6 Li}$ separation, but it has the substantial benefit of having a source of low cost isotope-selective photons from mercury vapor lamps for photoenrichment processes. It also benefits from having a significantly larger mass difference between the target isotope and most other isotopes, as large as six atomic mass units, as shown in Table \ref{tab:HgIsotopes}. This property would help significantly with mass-based separation methods including centrifugation, implemented either with conventional technologies or with plasma centrifuges.  

\subsubsection{Mercury Compatible Materials}

Prior work has already explored the range of materials with high temperature compatibility with mercury \cite{furman1971design,yamamoto1994experimental,weeks1964liquidus,rosenblum1968mechanism} and liquid lithium \cite{moynihan2024characterization, jun2020corrosion}.  

There is very limited information in the academic literature on $\mathrm{LiHg}$ alloys aside from studies on the phase diagram of this system \cite{hirayama_metals_1986}. If it is utilized, substantial work will be needed to better characterize compatibility with structural materials as well as to characterize thermophysical properties for this alloy that will impact blanket design. Use of this alloy is not required for implementation of the approach described here, since multiplier layers can be designed to only use mercury: as shown in Figure~\ref{fig:blanketscan} below a channel thickness of \SI{85}{cm}, at $100\mathrm{at}\%$ mercury the system still achieves TBR $>1.2$ with significant gold production.  Alternatively, more specialized blanket configurations can use separate channels for the $\ce{Li}$ and $\ce{Hg}$ in a multiplier layer so that the structural materials only have to be compatible with either material in each respective channel.

\subsubsection{Regulatory and Safety Challenges for Mercury} \label{sec:regsafety}


Given that large quantities of mercury will be used in these systems, this material will need to be regulated appropriately to avoid safety hazards or environmental release. 

The use of mercury is governed by the Minamata Convention on Mercury~\cite{MinamataConvention2024}, which (i) bans new primary mercury mining and phases out all remaining primary mines within fifteen years and (ii) regulates mercury use in manufacturing processes under Article 5.

Only preexisting uses of mercury are specifically restricted by the convention, and a mechanism exists for usage to be allowed for new processes that have environmental and health benefits \cite{MinamataConvention2024}. Given the significant benefits from large-scale adoption of fusion as well as phaseout of legacy gold mining processes, fusion transmutation appears to be a strong candidate for an acceptable new process. 

Since the process described here permanently transmutes mercury into a valuable material, it is possible that fusion transmutation could be considered as a form of waste disposal.  While early plants will be highly incentivized to specifically transmute $\ce{^{198}Hg}$, we note that the isotopes with higher neutron number can also in the long term be transmuted to $\ce{^{197}Au}$, though a large global transmutation capacity will be needed for this to be practical.  

 In the near term, the U.S.\ Department of Energy maintains a stockpile of $\sim 1700$ metric tons of mercury for NNSA purposes, and a recent estimate of annual mercury generation, primarily from ore processing in Nevada, is 130 metric tons per year \cite{USDOE2024Mercury}.
 In addition, the U.S.\ Department of Defense maintains a stockpile of nearly \num{5000} metric tons of material \cite{JMC2010Mercury}.
 The EU also has \num{6000} tons of mercury currently and expects to need to dispose of \num{11000} tons over the next 40 years \cite{EU2017Mercury,UBA2014Mercury}.
 As such, even with no change in existing processes, \num{14000} metric tons of mercury could be made available for processing and isotope removal in the next ten years of fusion development, corresponding to \num{1400} tons of \ce{^{198}Hg} and about the same mass of \ce{^{197}Au}, with a current market value of $\sim \$140\mathrm{B}$.  Depending on the progress made in reducing initial mercury inventories as described in Section \ref{inventorysubsection}, this material could be enough for tens to hundreds of gigawatts of fusion deployment.  

Over longer timelines additional feedstock mercury would be needed.
Mining of \ce{Hg} is far easier than mining of \ce{Au}, both because of far higher crustal abundance and because mercury ores form high concentration deposits.
Gold ores have concentrations of \ce{Au} of only 1\textendash100 ppm, requiring processing of massive streams of material for only a small total product quantity \cite{walshe_gold_2009}.  In contrast, $\mathrm{Hg}$-rich ores (e.g.\ cinnabar) have orders of magnitude higher concentration of $\mathrm{Hg}$ (several percent), and the material can be recovered through simple baking and distillation of the ore \cite{rytuba_mercury_2003}. Given the environmental impacts of gold mining and the benefits of fusion energy displacing other energy resources, regulators will need to carefully consider the benefits of maintaining well-regulated mercury mining to sustain transmutation.  

Use of mercury in the fusion system carries safety and environmental risks that power plant designers and operators must account for.  Robust controls are needed to ensure that any mercury feedstock is sourced from either waste or responsibly mined resources, and that environmental contamination is prevented along the entire supply chain for this material.  At the power plant itself, we expect that mercury will only be handled in elemental form or possibly alloyed with lithium, mitigating the risk of forming mercury compounds that can be significantly more toxic than the element itself.  In the power plant system, all materials will already be strictly controlled due to neutron activation and the presence of tritium, so there will be limited potential for mercury release in these systems with proper engineering controls.  Given the short term of decay for activation products in mercury discussed in Section \ref{sec:depletion}, we expect this material to have superior activated waste properties to lead, which produces significant amounts of long-lived radionuclides as well as radiotoxic \ce{^{210}Po} \cite{Pampin2006,Mertens2019}.  We also note that our proposal to use mercury for fusion is not unique, as it has also been proposed for use as a working fluid in pumping systems \cite{Giegerich2014}, as well as for application in COLEX/ICOMAX plants for lithium isotope separation \cite{Ward2025}.  Ultimately, the fusion industry will need to ensure risks are properly mitigated while maximizing the benefits of fusion energy for gold production and clean energy generation.

\subsection{Impact on Fusion Economics}

The consequence of this work is that the value of outputs of every D-T fusion power plant is now around twice what they were expected to be, assuming the current price of gold.  This will dramatically increase the amount of investment in fusion development and accelerate the deployment of economically viable fusion energy at scale.  

\begin{figure}[htbp]
  \centering
  \includegraphics[width=0.8\textwidth]{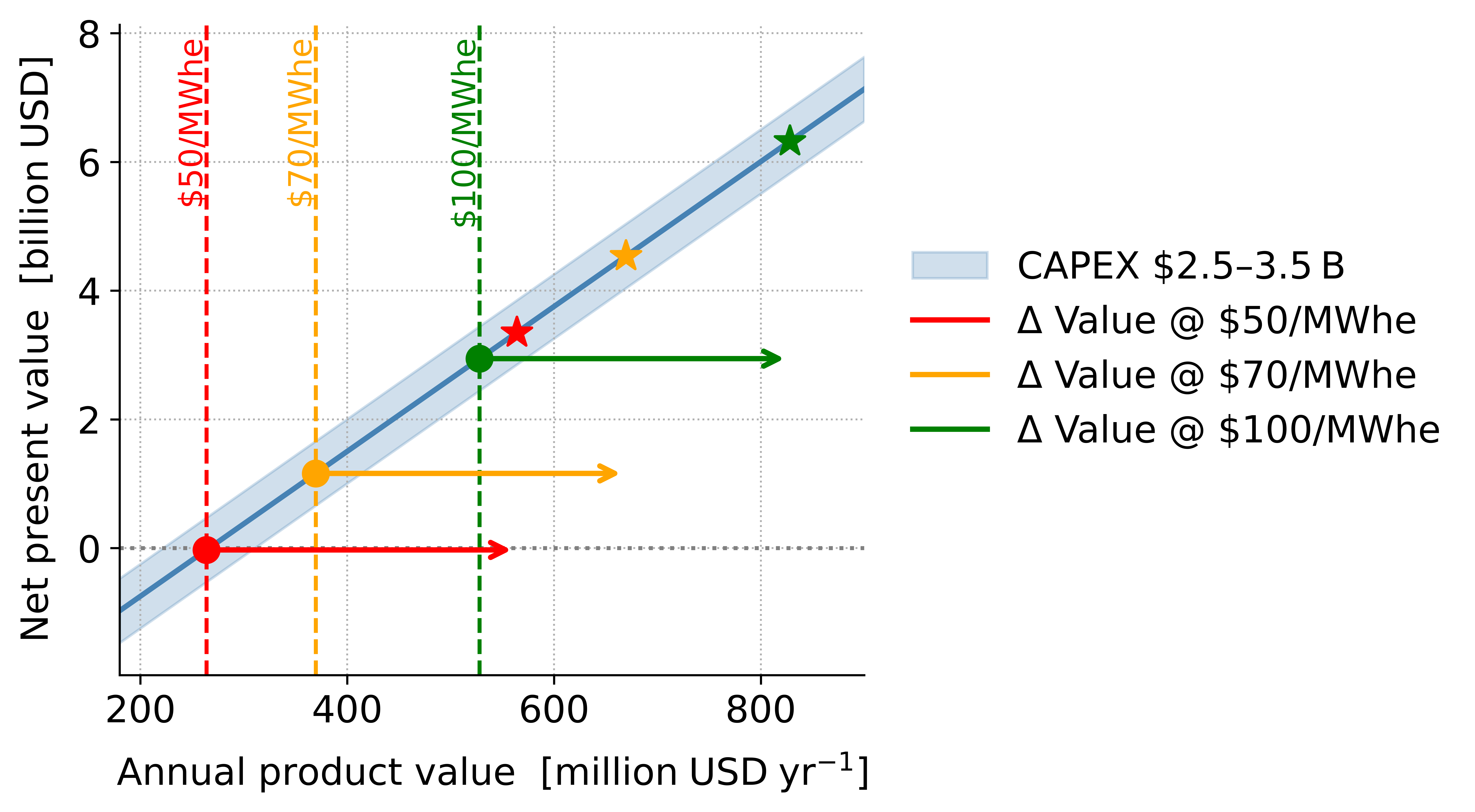}
  \caption{Net present value for a \$3B, $1500\,\mathrm{MW_{th}}$ fusion power plant operating at $100\%$ duty factor with $40\%$ conversion efficiency, and selling electricity for a range of three different assumed prices. The colored arrows show the increase in value from implementing gold production and the colored stars give the new net present value of each case.}
  \label{fig:npvfigure}
\end{figure}

We can gauge the overall benefit of this improvement by considering the effect on the net present value (NPV)---the discounted sum of all future cash flows minus the upfront capital cost---of an example fusion power plant.
Because investors require a positive NPV to commit capital to a project, this metric determines which markets fusion can enter.
The case shown in Figure~\ref{fig:npvfigure} illustrates the NPV of a $1500\,\mathrm{MW_{th}}$ fusion power plant with a capital expenditure (CAPEX) of \$3B and a discount rate of $8\%$ operating at a capacity factor of $100\%$ with electrical conversion efficiency of $40\%$ and a lifetime of 30 years.
Vertical dotted lines are shown for different electricity sale price assumptions, starting from the $\$50/\mathrm{MWh_e}$ target of \cite{Handley2021EarlyFusionMarkets} and including higher values which might be relevant to earlier but potentially more limited markets for fusion energy.  The shaded blue region represents an uncertainty bound for the plant CAPEX, placed between \$2.5--3.5B.
For each of these cases, the initial point represents the NPV for each initial annual product value and the stars represent the final net present value after including gold production, where the cost of gold is assumed to be $\$100\mathrm{k/kg}$, near the current value, and the plant is assumed to generate \SI{3}{t/yr} of gold.
With only electricity as the product, the NPV of the power plant is almost exactly at breakeven at $\$50/\mathrm{MWh_e}$; when the value of gold produced (here assumed to be $\$300\mathrm{M/yr}$) is included, the NPV is well over three billion dollars.

While there may be some increases to power plant CAPEX due to additional requirements for \ce{Au} generation, the fact that a multiplier layer and blanket are already needed in the system means that these additional costs will likely be modest.  An important observation from Figure~\ref{fig:npvfigure} is that the net present value does not simply double with a doubling of annual product value but can go from near zero to billions of dollars per plant.  The consequence is that many lower-value electricity markets that may not have previously been profitable could now be accessible to fusion energy, dramatically expanding the market opportunity for fusion applications.    

One of the most important properties of gold as a product material is that the market for this material is not saturated by a small number of fusion devices.
According to the U.S.\ Geological Survey, the total world production of gold in 2024 was 3300~tons~\cite{USGS_MCS_2025}.
Assuming for simplicity \SI{2}{t/GW_{th}/yr} of gold production, this approach subsidizes the deployment of \SI{1.65}{TW_{th}} of fusion power before matching the current mining capacity for gold.
The subsidy provided by gold production will continue well after this amount of fusion deployment, but with a lower value if the price of gold decreases as a result of increased supply.  

Beyond the near term impacts enabling the scaling of fusion, this work marks the start of the new field of industrial synthesis of valuable elements through transmutation.

\section{Conclusion} \label{sec5}


In this work, we have shown a practical pathway for large-scale, fusion-driven transmutation of mercury into gold, opening a new revenue stream for future D-T power plants while preserving electricity generation and tritium self-sufficiency. Monte Carlo transport calculations show that neutrons produced in a tokamak power plant can convert the abundant isotope $^{198}\mathrm{Hg}$ to stable $^{197}\mathrm{Au}$ via the (n,2n) channel, yielding several tonnes of gold per plant-year without compromising the tritium breeding ratio.

Beyond its symbolic significance - solving the alchemist's age-old quest using 21st-century nuclear engineering - the concept redefines fusion’s mission: from a stand-alone power technology to an integrated engine for valuable commodities and clean energy.

First, on the economic front, the sale of transmuted gold generates a revenue stream large enough to offset, and in many scenarios fully recover, capital and operating costs. This supplementary income sharply lowers the effective cost of the electricity produced and strengthens the business case for early deployment.

Second, because the economic breakeven point therefore occurs at markedly lower electrical output, the engineering demands on multiple subsystems are eased and the design space for fusion power plants is widened.

Third, from a materials and neutronics perspective, even in the absence of transmutation, $^{198}\mathrm{Hg}$ rivals or surpasses classical multipliers such as $^{7}\mathrm{Li}$, Be/FLiBe, or Pb in neutron multiplication, safety, and chemical reactivity.

Realizing this potential will require a coordinated R\&D programme in fusion transmutation engineering. Priorities include scalable enrichment of $^{198}\mathrm{Hg}$ and effort to minimize blanket inventory and waste streams; development of structural materials that tolerate liquid mercury at reactor-relevant temperatures and neutron fluences; and integrated design studies for magnetic-confinement, inertial-confinement, and magneto-inertial fusion blankets.

In summary, fusion-driven transmutation of $^{198}\mathrm{Hg}$ into gold transforms fusion energy from a stand-alone power technology into a multi-product industrial platform, dramatically strengthening its economic and societal value proposition. With focused effort on the technology gaps identified above, the approach described here could accelerate the commercial deployment of fusion power and, in doing so, turn an ancient aspiration into a reality. The goal of classical alchemy is now achievable through practical engineered solutions.

\section*{Acknowledgements}
\addcontentsline{toc}{section}{Acknowledgements}

The authors thank Jacob Schwartz, Ian Wojtowicz, Dan Brunner, Kyle Schiller, and Jehan Azad for helpful discussions and review of this manuscript. 

\section*{Data Availability Statement}
\addcontentsline{toc}{section}{Data Availability Statement}

The data presented in this study will be made openly available upon publication.

\appendix

\section{Other Reactions of Interest} \label{sec:otherreactions}

Though less economically interesting due to both lower price and much smaller market size, we also note that the methods described here can be applied to production of other precious metals.  While many pathways exist, we provide here a select few reactions for which the product material is high value, and there is a significant difference in cost between the feedstock material and the product material.  

 The same approach as described above can be used to produce palladium from silver via the reactions \cite{Augustyniak1975AgReactions, Luo2009Ag14MeV, kaeriAg106}

\begin{flushleft}

\[
\begin{aligned}
{}^{107}_{47}\mathrm{Ag}+n
&\xrightarrow[\;\sigma_{g}\;]{(n,2n)}
{}^{106\mathrm{g}}_{47}\mathrm{Ag}
\\[6pt]
{}^{107}_{47}\mathrm{Ag}+n
&\xrightarrow[\;\sigma_{m}\;]{(n,2n)}
{}^{106\mathrm{m}}_{47}\mathrm{Ag}
\\[10pt]
{}^{106\mathrm{g}}_{47}\mathrm{Ag}
&\xrightarrow[\;T_{1/2}=23.96~\text{min}\;]{\beta^{+}/\mathrm{EC}}
\begin{cases}
{}^{106}_{46}\mathrm{Pd} & (99.5\%)\\[4pt]
{}^{106}_{48}\mathrm{Cd} & (0.5\%)
\end{cases}
\\[10pt]
\textrm{and} \quad {}^{106\mathrm{m}}_{47}\mathrm{Ag}
&\xrightarrow[\;T_{1/2}=8.28~\text{d}\;]{\mathrm{EC}}
{}^{106}_{46}\mathrm{Pd}\;(100\%).
\end{aligned}
\]
\end{flushleft}
This product has the challenge of having a large portion of natural abundance of an isotope of the feedstock element which would generate long-lived radioactive waste if not separated out from the feedstock material.  Specifically, \ce{^109Ag} represents $48.2\%$ of naturally occurring \ce{Ag}, and $(n, 2n)$ reactions on this material produce both the ground state $^{108\mathrm{g}}\mathrm{Ag}$ which has a half life of only 2.38~minutes, but also metastable $^{108\mathrm{m}}\mathrm{Ag}$ which has a half life of 418~years \cite{KAERI_Ag108_2025}. 

There is also a pathway to produce osmium from rhenium through the reactions~\cite{kaeriAtomRe186, MIRDspecs_Re186m_2024}  
\[
\begin{aligned}
{}^{187}_{75}\mathrm{Re} + n
&\xrightarrow[\;\sigma_{g}\;]{(n,2n)}
{}^{186\mathrm{g}}_{75}\mathrm{Re}
\\[6pt]
{}^{187}_{75}\mathrm{Re} + n
&\xrightarrow[\;\sigma_{m}\;]{(n,2n)}
{}^{186\mathrm{m}}_{75}\mathrm{Re}
\\[10pt]
{}^{186\mathrm{m}}_{75}\mathrm{Re}
&\xrightarrow[\;T_{1/2}=2.0\times10^{5}\,\text{y}\;]{\mathrm{IT}}
{}^{186\mathrm{g}}_{75}\mathrm{Re}
\\[10pt]
\textrm{and}\quad {}^{186\mathrm{g}}_{75}\mathrm{Re}
&\xrightarrow[\;T_{1/2}=3.7186\,\text{d}\;]{\beta^-}
\begin{cases}
{}^{186}_{76}\mathrm{Os} & (92.53\%)\\[6pt]
{}^{186}_{74}\mathrm{W}  & (7.47\%).
\end{cases}
\end{aligned}
\]
The above two reactions represent options in which there is a large ratio of product cost to material cost, indicating the potential for economic feasibility.  

While not particularly interesting economically due to the rather low price ($\sim \$10^3\mathrm{/kg}$ currently \cite{LBMA_Silver_Price_2025}), we note that silver may also be produced from the relatively abundant $^{110}\mathrm{Cd}$, which has $12.5\%$ natural abundance \cite{kaeri_cd110}. After $(n,2n)$, the \ce{^{109}Cd} produced decays through electron capture with a half life of $464.6$ days to stable $^{109}\mathrm{Ag}$ \cite{kaeri_cd109}.  Aside from poor economics, this process is impractical due to the very long half-life of the intermediate material.  However, we note it here due to its historical significance as a scalable solution to the traditional alchemical pursuit of \textit{argyropoeia}, or artificial production of silver.  

Aside from production of precious metals, the same approach proposed here can also be applied to production of valuable stable isotopes and radioisotopes.  Future work will discuss potential applications for medical isotopes and nuclear batteries. 

\section{Reactions Producing \texorpdfstring{$^{197}$}{197}Au from JENDL-5}
\label{sec:Auxsections}
\begin{figure}[htbp]
  \centering
  \centering
  \includegraphics[width=0.85\textwidth]{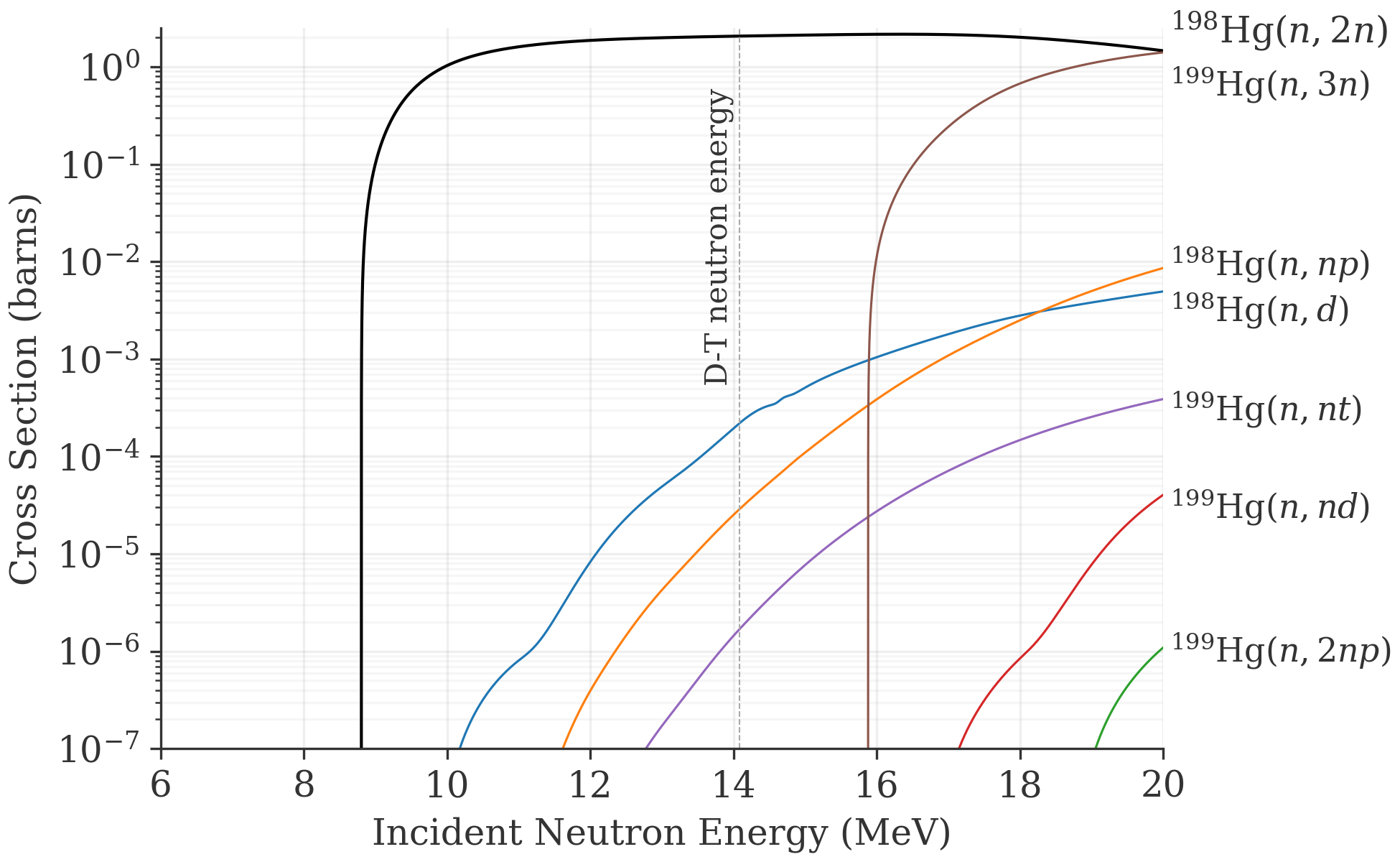}
  \caption{Cross sections for $^{197}$Au producing reactions shown in Table \ref{tab:AuReactions} across relevant energy ranges.}
  \label{fig:Au197_cross_sectionsA}
\end{figure}

\begin{figure}[!htbp]
  \centering
  \centering
  \includegraphics[width=0.65\textwidth]{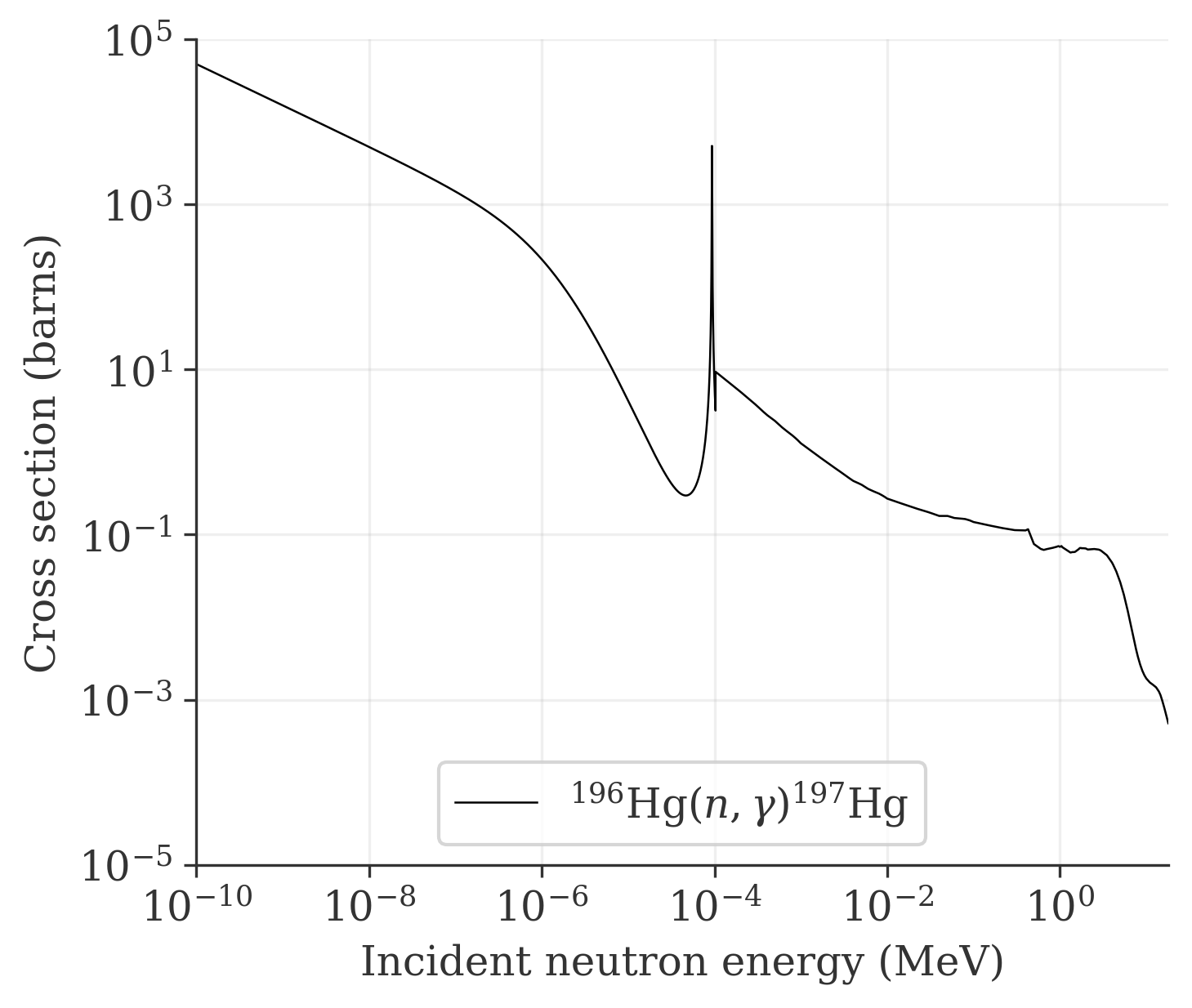}
  \caption{Cross section for \ce{^{196}Hg(n,\gamma)^{197}Hg}. The product subsequently decays to \ce{^{197}Au}.
  }
  \label{fig:Au197_cross_sectionsB}
\end{figure}

This appendix describes the reactions that lead to the production of \ce{^197Au} which are included in the OpenMC simulations.
Table~\ref{tab:AuReactions} lists the initial nuclides, the reaction, the main products, and the fraction of Au production due to each pathway.
Figure~\ref{fig:Au197_cross_sectionsA} shows the cross sections of the MeV-energy $(n,*)$ reactions, and Figure~\ref{fig:Au197_cross_sectionsB} shows the cross section of \ce{^{196}Hg(n,\gamma)^{197}Hg}, which increases at lower energy. 

\begin{table}[h]
\centering
\begin{tabular}{@{}llll@{}}
\hline
\textbf{Initial nuclide} & \textbf{Reaction} & \textbf{Main product(s)} & \textbf{$\%$ Contribution to Au Production} \\ \hline
$^{198}$\text{Hg} & $(n,2n)$      & $^{197}$\text{Hg} $\;\xrightarrow{\beta^-}\;$ $^{197}$\text{Au} & 99.9865140671\% \\
$^{196}$\text{Hg} & $(n,\gamma)$  & $^{197}$\text{Hg} $\;\xrightarrow{\beta^-}\;$ $^{197}$\text{Au} & 0.0030773036\% \\
$^{198}$\text{Hg} & $(n,d)$       & $^{197}$\text{Au} & 0.0090771331\% \\
$^{198}$\text{Hg} & $(n,np)$      & $^{197}$\text{Au} & 0.0013286591\% \\
$^{199}$\text{Hg} & $(n,t)$       & $^{197}$\text{Au} & 0.0000017002\% \\
$^{199}$\text{Hg} & $(n,nd)$      & $^{197}$\text{Au} & 0.0000000000\% \\
$^{199}$\text{Hg} & $(n,2np)$     & $^{197}$\text{Au} & 0.0000000000\% \\
$^{199}$\text{Hg} & $(n,3n)$      & $^{197}$\text{Hg} $\;\xrightarrow{\beta^-}\;$ $^{197}$\text{Au} & 0.0000011369\% \\ \hline
\end{tabular}
\caption{Full list of $^{197}$Au producing reactions included in neutronics simulations. Contributing percentages taken from the case of \SI{210}{mm} channel thickness with $85\%$ mercury concentration. }
\label{tab:AuReactions}

\end{table}

\clearpage

\bibliography{bibliography.bib}

\end{document}